\documentclass[trackchanges]{aastex701}

\usepackage{amsmath}
\usepackage{amssymb}
\usepackage{tipa}
\usepackage{CJKutf8}
\usepackage{wasysym}
\usepackage{combelow}
\usepackage{graphicx}
\usepackage{subcaption}

\begin{document}

\title{Probabilistic Spectral Reconstruction of Trans-Neptunian Objects from Sparse Photometry: A Framework for Taxonomy, Survey Optimization, and Outlier Detection}

\author[0000-0001-7737-6784]{Hsing~Wen~Lin (\begin{CJK*}{UTF8}{bkai}
林省文\end{CJK*})}
\affiliation{Department of Physics, University of Michigan, Ann Arbor, MI 48109, USA}
\affiliation{Michigan Institute for Data and AI in Society, University of Michigan, Ann Arbor, MI 48109, USA}
\email{hsingwel@umich.edu}

\author[0000-0002-2486-1118]{Larissa Markwardt}
\affiliation{University of Auckland, Science Centre 303, 38 Princes St, Auckland Central, Auckland, 1010, New Zealand}
\email{larissa.markwardt@auckland.ac.nz}

\author[0000-0003-4827-5049]{Kevin J. Napier} 
\affiliation{Center for Astrophysics $\mid$ Harvard \& Smithsonian, Cambridge, MA 02138, USA}
\affiliation{Michigan Institute for Data and AI in Society, University of Michigan, Ann Arbor, MI 48109, USA}
\affiliation{Department of Physics, University of Michigan, Ann Arbor, MI 48109, USA}
\email{kjnapier@umich.edu}

\author[0000-0002-8167-1767]{Fred C. Adams}
\affiliation{Department of Physics, University of Michigan, Ann Arbor, MI 48109, USA}
\affiliation{Department of Astronomy, University of Michigan, Ann Arbor, MI 48109, USA}
\email{fca@umich.edu}

\author[0000-0002-1226-3305]{Renu Malhotra}
\affiliation{Lunar and Planetary Laboratory, The University of Arizona, Tucson, AZ 85721, USA
Tucson, AZ 85721, USA}
\email{malhotra@arizona.edu}

\author[0000-0001-6942-2736]{David W. Gerdes}
\affiliation{Department of Physics, Case Western Reserve University, Cleveland, OH 44106, USA}
\affiliation{Department of Physics, University of Michigan, Ann Arbor, MI 48109, USA}
\affiliation{Department of Astronomy, University of Michigan, Ann Arbor, MI 48109, USA}
\email{dwg50@case.edu}

\correspondingauthor{Hsing~Wen~Lin}
\email{hsingwel@umich.edu}

\begin{abstract}

Near-infrared (near-IR) spectroscopy provides critical constraints on the surface composition of trans-Neptunian objects (TNOs), but spectroscopic observations remain limited compared to broadband photometry. We develop a probabilistic latent-space framework to quantify how much spectral information is retained in sparse photometric measurements. Using a principal component representation trained on a sample of near-IR spectra, we model the spectral manifold of TNOs and perform Bayesian inference in this reduced space to reconstruct full spectra from photometry while propagating measurement uncertainties. Leave-one-out cross-validation demonstrates that the dominant modes of spectral variability are low-dimensional: approximately 4 to 5 principal components capture the structure relevant for taxonomic classification, while 8–10 components improve spectral reconstruction fidelity and uncertainty calibration. For most objects, the reconstructed spectra achieve empirical credible-interval coverage of $\sim95\%$ across wavelength. These results suggest that the diversity of near-IR spectral shapes in the current sample is governed by structured, correlated surface processes rather than stochastic variation. Practically, we apply this framework to survey optimization, quantifying the information content of JWST/NIRCam filters to identify optimally informative configurations (e.g., F090W, F115W, F410M, F460M) for future observations for TNO taxonomy. Additionally, we demonstrate the pipeline's capability to detect and reconstruct rare spectral types, such as the peculiar Neptune Trojans 2006~RJ$_{103}$ and 2011~SO$_{277}$, by allowing constraining photometry to select low-probability intermediate models from the continuous topological manifold. Ultimately, this framework bridges the gap between sparse photometry and spectroscopy, providing a statistically rigorous tool to map the compositional structure of minor planets in upcoming large-scale surveys.

\end{abstract}

\keywords{Trans-Neptunian objects (1705) --- Neptune trojans (1097) --- Broadband photometry (184) ---Infrared Spectroscopy (2285) --- Multi-color photometry(1077)}

\section{Introduction} 
\label{sec:intro}
Trans-Neptunian Objects (TNOs) are the primordial small bodies of our Solar System. Their orbital distributions and surface colors serve as fundamental tracers for modeling the formation and evolutionary history of the solar system \citep[e.g.,][]{Tegler1998, Nice, Nesvorny2020}. While TNOs exhibit a well-known, wide color distribution \citep{Bernardinelli2025, Fraser2023, Pike2017, Schwamb2019, Peixinho2012, Peixinho2015} that is directly linked to their surface composition, spectroscopic observations are essential to truly constrain their chemical components. However, the faintness and large heliocentric distances of TNOs make direct spectroscopy difficult; consequently, photometry remains the most efficient method for studying their surface properties. While their optical spectra are often featureless, the diagnostic near-infrared (near-IR) wavelengths are difficult to observe from the ground, requiring space-based facilities like the James Webb Space Telescope (JWST).

Recent JWST/NIRSpec observations have revealed three principal TNO surface types: a Water-type, dominated by water-ice absorption; a CO$_2$-type, characterized by strong CO$_2$ bands with a distinct double-dip near 4.26 $\mu$m; and an Organics-type, showing aliphatic features between 3.3–3.6 $\mu$m \citep{disco2024, Holler2025}. The Organics-type can be further subdivided into methanol-rich and methanol-poor subtypes, with the latter commonly found among Cold Classical Kuiper Belt Objects \citep{Brunetto2025}. These spectral distinctions likely reflect the condensation fronts of key volatiles, preserving a record of the Solar System’s primordial compositional gradient.

Although JWST has enabled systematic spectroscopy of these distant bodies, such observations remain time-consuming and expensive. Currently, low-resolution near-IR spectra have been obtained for only $\sim$75 objects. Furthermore, spectroscopic targets are biased toward brighter, larger, and closer objects. To effectively study the surface properties of the TNO population across a broad range of sizes and distances, deep near-IR photometric surveys remain the most effective strategy. The JWST Cycle 3 program \#6064 \citep{jwst6064} and ongoing JWST Cycle 4 program \#7248 \citep{jwst7248} are examples of such efforts, and this spectrophotometry technique has also been applied to study the small satellites of Uranus and Neptune \citep{Belyakov2024}.

Using photometry as a proxy for spectroscopy is an established concept in minor planet studies. This approach has been successfully applied to Main Belt and Near-Earth Asteroids \citep{DeMeo2013} based on the foundational taxonomy of asteroid spectra \citep{Bus2002a, Bus2002b, DeMeo2009}. By approximating the spectral type of thousands of asteroids, these studies have enabled compositional mapping of the asteroid belt that critically constrains solar system evolution models \citep{DeMeo2014}.

The spectral taxonomy of asteroids defined by \citet{DeMeo2009} and \citet{DeMeo2013}, as well as the TNO taxonomy in optical colors \citep{Barucci2005} and near-IR spectra \citep{disco2024}, relies heavily on Principal Component Analysis (PCA). More recently, several studies have applied modern machine learning techniques, including neural networks and Transformer architectures, to classify asteroid spectra \citep{delbo2026, Tang2025, Luo2024, Penttila2021}. These approaches are primarily discriminative models operating on existing high-resolution spectral data. \citet{Penttila2022} extended this framework to spectrophotometric measurements, but similarly focused on forward taxonomy classification. In contrast, our work addresses the inverse problem: inferring continuous spectral structure probabilistically from sparse photometric inputs, enabling characterization of faint objects prior to spectroscopic follow-up.

Here, we propose an alternative probabilistic framework for TNO taxonomy. Unlike previous PCA-based studies that emphasize dimensionality reduction or deterministic classification, we treat the latent representation as a generative prior within a fully Bayesian framework, enabling posterior sampling and empirical calibration of spectral reconstructions from sparse photometry. This approach parallels forward-modeling and simulation-based inference techniques developed in other areas of astronomy, including likelihood-free posterior estimation in cosmology \citep[e.g.,][]{Alsing2018, Alsing2019, Jeffrey2021} and modern hierarchical Bayesian modeling practices \citep[e.g.,][]{Hogg2010, Ivezic2014}. By explicitly modeling the data-generating process, our method propagates photometric into the reconstructed spectral manifold and provides calibrated probabilistic constraints on surface composition. The structure of this paper is as follows: Section~\ref{sec:method} describes the conceptual basis and architecture of the reconstruction framework. Section~\ref{sec:perf} describes the procedure of model training, validation, and calibration. In Section~\ref{sec:app}, we present potential applications for finding the most informative filters configurations for future near-IR TNO photometry surveys, and spectral outlier identification. We discuss the mathematical basis of the framework's performance, and the limitations of our model in Section \ref{sec:disc}. Finally, a summary of this work and future development prospects are provided in Section~\ref{sec:summary}.

\section{Methodology} \label{sec:method}

\subsection{Bayesian framework}

The challenge of estimating a continuous spectrum from discrete photometric points can be mathematically framed as \textit{spectral super-resolution}. Originally pioneered in computer vision to reconstruct high-fidelity images from down-sampled inputs, super-resolution algorithms overcome the sampling limits of sensors and optics by using a learned ``prior'', a statistical understanding of valid high-resolution structures \citep{bashir2021comprehensive, bizhani2022reconstructing, huang2015single, tai2010super, wang2015deep, vandewalle2017registration}. Such techniques were also applied on astronomical images \citep{li2025benchmark, swanson2025super, Brout2019, bernardinelli2024photometry}. While traditionally applied to spatial domains, the underlying principle is directly applicable to the spectral domain of astronomical observations. In this context, multi-band photometry represents a low-resolution sampling of a celestial body's Spectral Energy Distribution (SED). The \textbf{prior} is the accumulated knowledge of the spectral morphology for a specific population of celestial bodies. The process of spectral reconstruction is thus well-described as a Bayesian inference problem:

\begin{equation}
\label{eq:bayesian_spectral}
\underbrace{p(S | \mathbf{m})}_{\text{Posterior}} \propto \underbrace{\mathcal{L}(\mathbf{m} | S)}_{\text{Likelihood}} \times \underbrace{p(S)}_{\text{Prior}},
\end{equation}
\noindent where $S$ represents the high-resolution\footnote{Here, ``high-resolution'' implies a spectral resolution significantly higher than that of the sparse photometric filters.} spectrum vector, and $\mathbf{m}$ represents the observed sparse \textbf{Photometry}. $p(S)$ is the \textbf{Prior}: implicitly defined by the PCA basis and the kernel density estimate over the latent coefficients, which together induce a probability distribution over physically plausible spectra. $\mathcal{L}(\mathbf{m} | S)$ is the \textbf{Likelihood}: the probability of observing photometry $\mathbf{m}$ given a true spectrum $S$. \textcolor{black}{Although explicitly defined here, we approximate this likelihood mapping via Simulation-Based Inference (see Section~\ref{sec:implicit_likelihood}).} Finally, $p(S | \mathbf{m})$ is the \textbf{Posterior}: the reconstructed spectrum given the photometric observations. The latent coefficients represent the underlying physical or statistical patterns of the spectra in a reduced-dimensional space. They act as a compressed blueprint from which the full, high-dimensional spectral features can be reconstructed. 

\subsection{The Architecture}
Following this super-resolution and Bayesian framework concept, we propose an architecture that operates as a physically constrained encoder-decoder pipeline designed to infer high-dimensional spectral information from low-dimensional photometric observables. The workflow proceeds in two phases: latent space definition and forward inference. 

First, we employ Principal Component Analysis (PCA, e.g. \citealp{Ivezic2014}) on a set of reference TNO spectra to define a low-dimensional latent space (the Prior). This space serves as a coordinate system where the eigenvectors represent fundamental variations in surface properties. During inference, the inverse PCA transform effectively acts as the ``decoder,'' providing a deterministic mapping from compressed latent variables back to full-resolution spectra.

\begin{figure}
\includegraphics[width=.5\columnwidth]{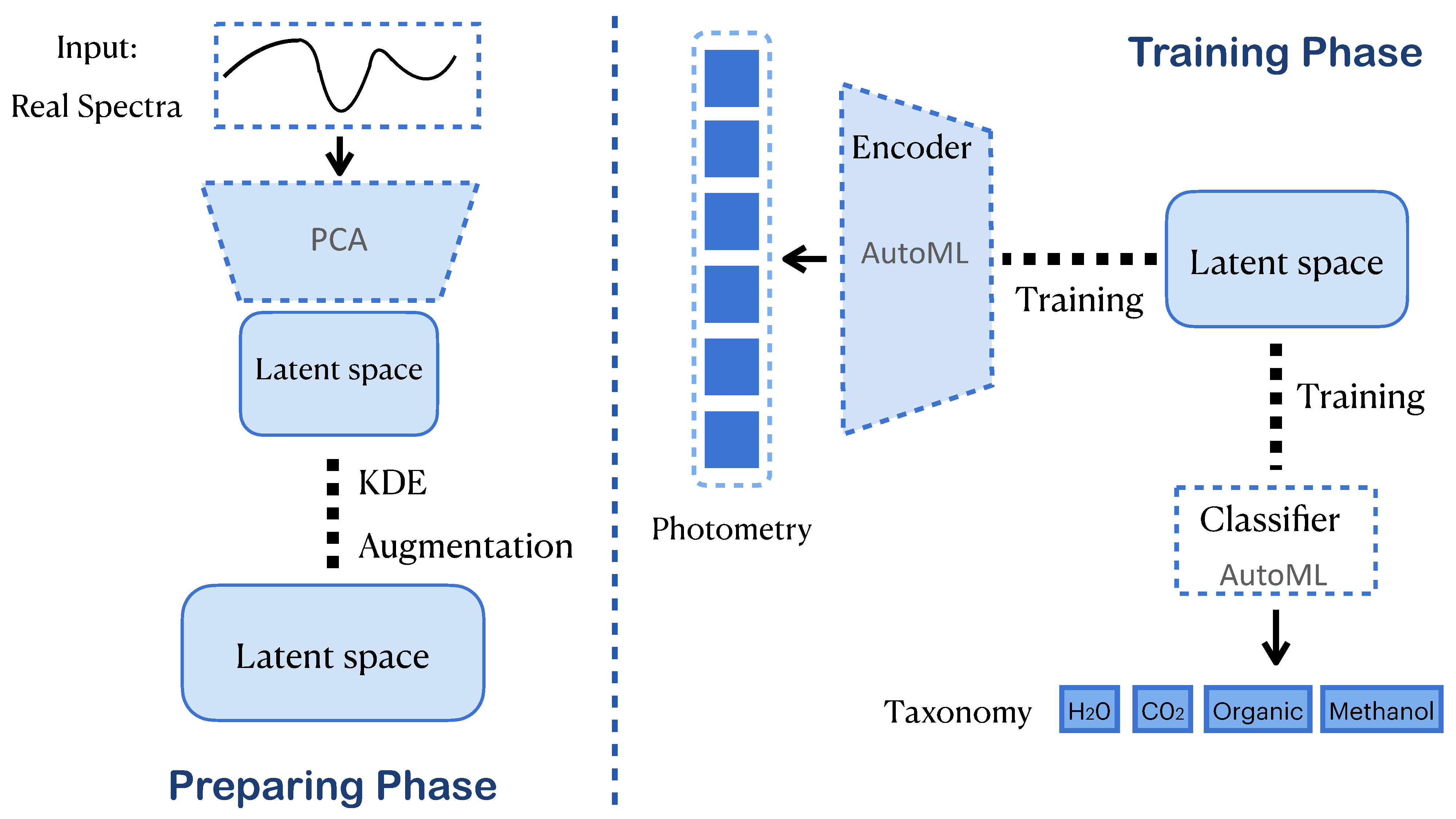}
\includegraphics[width=.5\columnwidth]{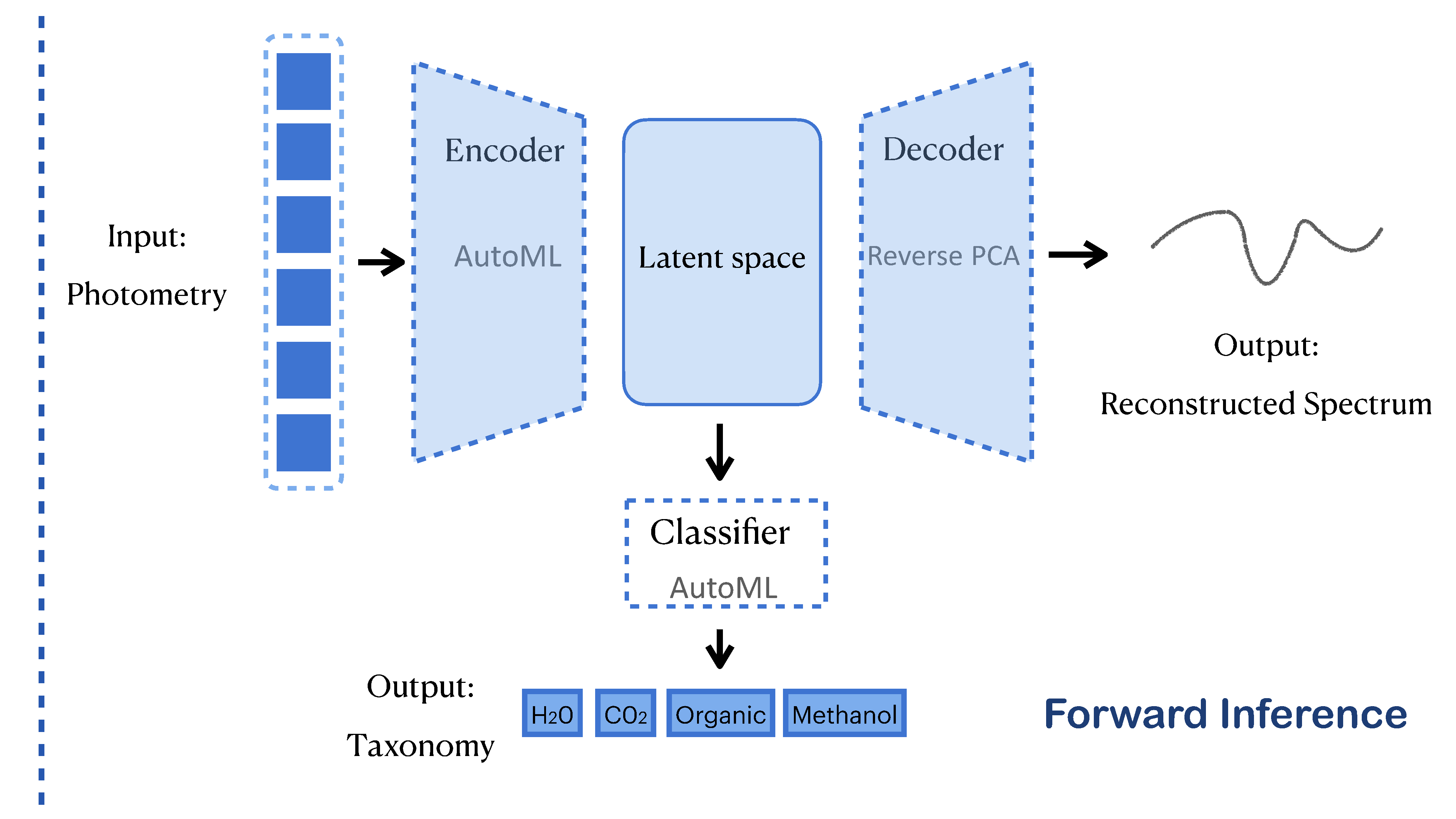}
\caption{The architecture and workflow of the TNO spectral reconstructor and classifier, divided into three operational phases. \textbf{(Left) Preparing Phase:} Observational seed spectra are compressed via Principal Component Analysis (PCA) to define the initial latent manifold, which is subsequently augmented using Kernel Density Estimation (KDE) to map a continuous probability distribution. \textbf{(Middle) Training Phase:} The augmented latent coordinates and their corresponding forward-modeled mock photometry are used to train the AutoML regressor (the Encoder) and the AutoML taxonomic classifier. \textbf{(Right) Forward Inference:} New sparse photometric observations are passed through the trained encoder to predict a probability distribution within the latent space. This predicted state is simultaneously passed to the classifier for taxonomic identification (H$_2$O, CO$_2$, Organic, Methanol) and projected through the inverse PCA basis (the Decoder) to generate a reconstructed spectrum.}
\label{fig:arch}
\end{figure}

Second, we train an Automated Machine Learning (AutoML) regressor, specifically an ensemble of gradient-boosted decision trees \citep[e.g.,][]{XGBoost}, to serve as the ``encoder.'' This model learns the non-linear mapping from sparse input photometry \textcolor{black}{(representing the observational constraints)} to the latent coefficients. During inference, the pipeline accepts new photometric observations, predicts their corresponding latent coordinates, and projects them through the inverse PCA transform to reconstruct the spectra. The overview of the architecture is shown in Figure~\ref{fig:arch}.

A critical design choice in our architecture is the decision to regress the photometric inputs onto the low-dimensional PCA coefficients rather than attempting to predict the full-resolution reflectance at every wavelength bin directly. Mapping a sparse input vector (e.g., 2--6 photometric points) to a high-dimensional output vector (e.g., hundreds of spectral bins) constitutes a severely ill-posed inverse problem \citep{Tarantola1987}, where small perturbations in the input can lead to chaotic, unphysical oscillations in the output. By targeting the PCA coefficients, we effectively reduce the mapping from hundreds of spectral bins to a small set of latent coefficients (typically $K \le 10$, \textcolor{black}{the choice of K will be discussed in Section~\ref{sec:k_selection}}), making the inverse problem numerically stable even for sparse photometry. Furthermore, the principal components act as a strong physical regularizer: because the reconstruction is restricted to linear combinations of dominant eigenvectors, the output spectrum inherits the covariance structure and smoothness characteristics of the training population. This approach prevents the machine learning model from ``hallucinating'' high-frequency noise or independent pixel-to-pixel variations that are physically impossible given the mineralogical constraints of the surface.

We utilized AutoML, specifically the \texttt{AutoGluon} framework \citep{agtabular}, to drive the photometric-to-latent-space regression. In many applications of deep learning to astrophysics, significant research effort is often diverted toward manual hyperparameter tuning and architectural engineering. By automating the model selection process, ensembling robust algorithms like Gradient Boosted Decision Trees (GBDTs) and neural networks, AutoML allows us to abstract away the implementation details. This strategy ensures that our performance metrics reflect the true information content of the photometric data rather than the skill in tuning a specific network.

\subsection{The Prior: Training Data}
\label{sec:prior}
Since the prior is the probability distribution of valid TNO spectra, we start with the TNO spectra from the DiSCo-TNO sample (\citealp{disco2024}, \dataset[DOI: 10.17909/r2zp-r280]{https://doi.org/10.17909/r2zp-r280}). The sample was intentionally restricted to the dominant taxonomic classes (e.g., the Water-type, the CO$_2$-type, and the Organics-type) while excluding minority spectral types and singular outliers, such as Haumea family members, Dwarf planets \citep{Emery2024}, Centaurs \citep{discocentaur2024}, blue binaries of cold classical Kuiper belt objects \citep{Wong2025PSJ}, and Neptune Trojans \citep{Markwardt2025}. Therefore, this prior is not universally applicable to all outer Solar System populations.

This exclusion is driven by the statistical requirements of Principal Component Analysis (PCA). In a manifold learning framework, including rare morphotypes in a small training sample would unreasonably skew the resulting eigenvectors, degrading the reconstruction quality for the vast majority of the population. By constructing a ``clean'' latent space that represents only the continuous physics of the dominant TNO population, we enforce a strong prior that maximizes stability for standard objects. 

Given the observational expense and the resulting scarcity of high-quality near-infrared TNO spectra discussed in Section~\ref{sec:intro}, our baseline dataset is inherently restricted. Therefore, a rigorously vetted sample of 51 TNO spectra was selected as the seed population for the probability distribution. 
To overcome the severe sparsity of the spectroscopic training set ($N \approx 50$), which is insufficient to train a robust machine learning regressor without overfitting, we employed a generative data augmentation strategy based on manifold learning. 

We first projected the seed spectra onto their principal components to obtain distributions of latent coefficients. We evaluated two methods for density estimation within this latent space: non-parametric Kernel Density Estimation (KDE) and parametric Gaussian Mixture Models (GMM). This choice represents a fundamental bias-variance tradeoff in manifold design: a GMM creates a highly restricted prior, ensuring that the model only generates secure, standard taxonomic spectra without hallucinating morphologies. On the other hand, KDE constructs a more diverse, continuous manifold, but with the risk of creating ``Frankenstein'' spectra, the physically improbable hybrid spectra, in the intermediate regions. However, the ability to generate such intermediate spectra is actually highly advantageous under this generative framework; we refer the reader to Section~\ref{sec:outliers} for details.

We found that fitting a multi-component GMM in a high-dimensional space ($K \ge 5$) with a seed population of only $\sim 50$ objects, particularly when certain spectral types are sparsely populated, introduces a severe risk of overfitting and parameter non-convergence. Restricting the latent space to fewer dimensions to accommodate a stable GMM would result in an unacceptable loss of fine spectral detail. Moreover, applying a GMM on the small dataset, it would effectively devolve the generative pipeline into a complex nearest-neighbor ``lookup table''.
Given these constraints, we opted for the non-parametric KDE, as it naturally and robustly smooths the probability density between discrete observations. \textcolor{black}{Specifically, we utilized a Gaussian kernel to ensure a smooth, continuous probability density across the latent space.} This allowed us to sample thousands of synthetic latent vectors, effectively interpolating between discrete real-world examples to create a continuous, smooth manifold.

The KDE bandwidth ($h$) controls the degree of smoothing applied to the latent density and therefore regulates the balance between local fidelity and global manifold smoothness. If the bandwidth is too small, the prior becomes highly multimodal and fragmented, leading to unstable regression behavior and poor predictive calibration. Conversely, excessively large bandwidth values over-smooth the latent distribution and generate synthetic spectra representing unrealistic combinations of taxonomic features. We experimented with several bandwidth values from 0.1 to 1 found that $h \sim 0.5$ provides stable predictive calibration while preserving physically plausible spectral morphologies and class separability in latent space. Results are modest perturbations around this value.

\begin{figure}

\includegraphics[width=1\columnwidth]{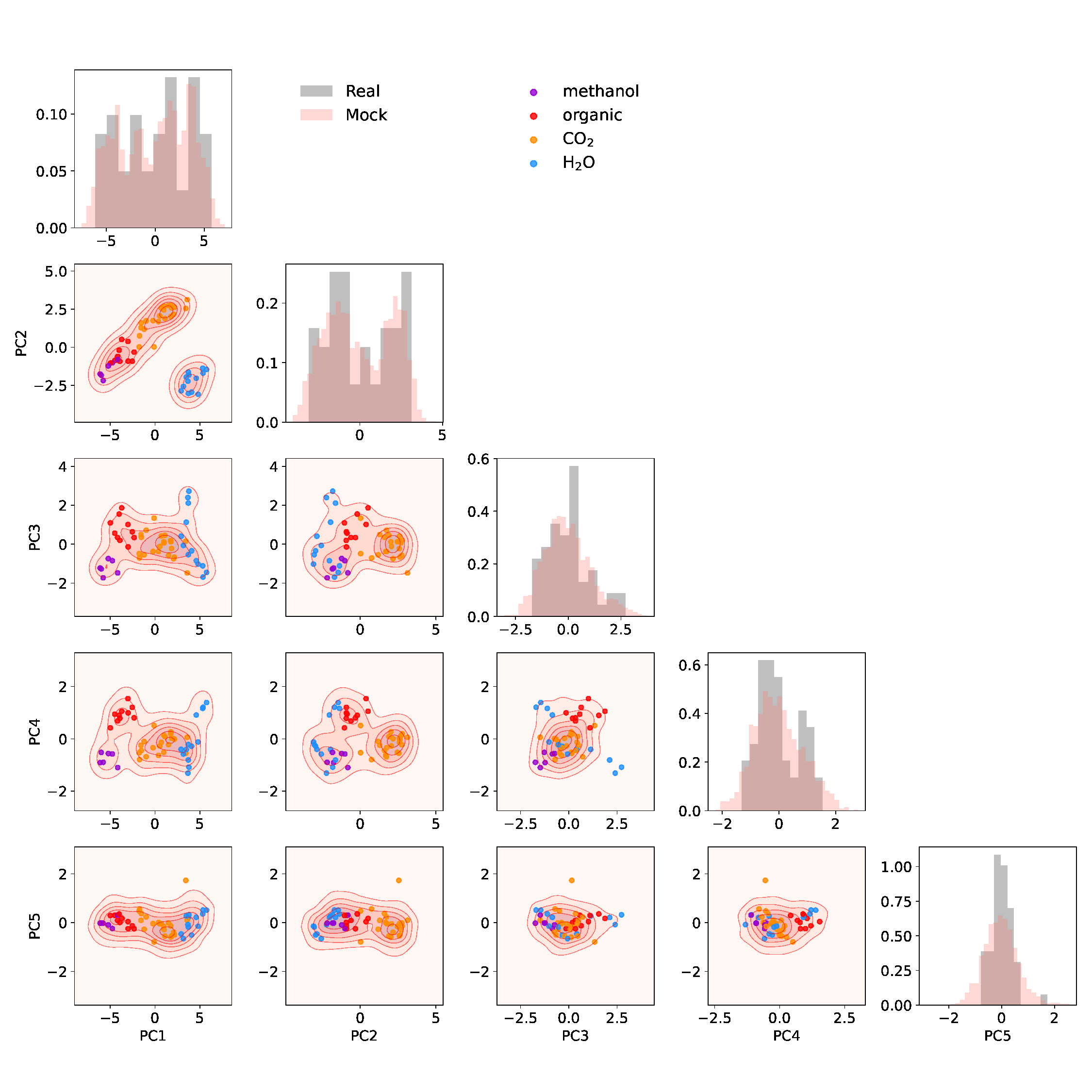}

\caption{An illustration of synthetic training data (mock) and the observational seed spectra distributions in a 5-dimensional manifolds.}

\label{fig:latent}

\end{figure}

Figure~\ref{fig:latent} illustrates the distribution of both the synthetic training data (mock) and the observational seed spectra (real) projected onto the 5-dimensional principal component latent space. The seed spectra are color-coded into four distinct compositional groups: \textbf{methanol (purple)}, representing methanol-rich organic spectra; \textbf{organic (red)}, representing methanol-poor organic spectra; \textbf{CO$_2$ (orange)}, corresponding to CO$_2$-rich spectra; and \textbf{H$_2$O (blue)}, indicating water-ice dominated spectra. As shown, the mock data generally trace the locus of the seed spectra while effectively filling the intermediate regions of the PC space. 

It is important to emphasize that the KDE augmentation is used solely to regularize the photometric-to-latent regression model and does not enter the evaluation stage. All performance metrics reported in Section~\ref{sec:perf}, including leave-one-out cross validation (LOOCV) reconstruction error, empirical coverage, and taxonomy classification are computed exclusively on the original 51 observed spectra. During each LOOCV iteration, the held-out real spectrum is removed from the seed set prior to density estimation and augmentation, ensuring that neither the latent density model nor the regressor has access to the validation target. The synthetic samples therefore serve only to provide a smooth approximation to the underlying spectral manifold and to stabilize regression training in the low-sample regime; they do not bias or contaminate the cross-validation assessment. We refer to Section~\ref{sec:LOOCV} for the details of the LOOCV process.

\textcolor{black}{\subsection{The Implicit Likelihood: Forward Modeling}
\label{sec:implicit_likelihood}}
To establish the mapping between observational data and the latent spectral space, \textcolor{black}{we employ a Simulation-Based Inference (SBI, or likelihood-free inference) approach rather than defining an explicit, analytical likelihood function (such as a standard $\chi^2$ comparison). First,} we generated a corresponding set of synthetic photometry for each augmented spectrum via forward modeling. We convolved the spectra with the effective throughput curves of the target instrument, specifically the JWST NIRCam filter set, to calculate the theoretical reflectance in each band\footnote{While we use reflectance (the observed spectrum divided by the solar spectrum), one can also use magnitudes or flux densities where the solar spectrum is not removed. Since there is no information gain or loss between reflectance and flux, the final results will be the same.}.

To \textcolor{black}{implicitly construct this likelihood mapping and} ensure the machine learning model remains robust to observational uncertainties, we injected Gaussian noise at a relative reflectance uncertainty level of 20\% into the synthetic photometry. In reality, photometric uncertainties depend on the specific filters and exposure times, and therefore should vary between the chosen bands. The 20\% uncertainty used here serves as a conservative ``worst-case'' scenario, corresponding to borderline $S/N \approx 5$ detections. This perturbation forces the regressor to learn the underlying \textcolor{black}{probabilistic} correlations rather than overfitting to precise, noise-free mathematical determinism. This process results in a comprehensive training database of noisy photometry inputs paired with ground truth PCA targets, effectively simulating the real-world conditions of a deep photometric survey \textcolor{black}{and representing our empirical likelihood distribution.}

\textcolor{black}{Within our evaluation framework, it is important to note that this entire forward-modeling and likelihood-simulation process is performed strictly on the continuous manifold built from the $N-1$ training objects (see Section~\ref{sec:LOOCV}). Once the global mapping is learned by the AutoML regressor, the measured photometry of the single held-out object is then passed through the pipeline to compute its specific posterior.}
\subsection{The Posterior: Reconstructed Spectra}
The final reconstruction step acts as the generative decoder of our pipeline, synthesizing the full spectra from the predicted latent variables. By applying the inverse PCA transform to the coefficients output by the AutoML regressor, we recover a high-resolution reflectance spectrum. Because this reconstruction is formed by a linear combination of the principal eigenvectors, it inherits the physical smoothness and characteristic absorption features inherent to the training population. The reconstruction acts as a regularized estimator: observational scatter in individual bands does not propagate into high-frequency spectral fluctuations, because the solution is constrained to the learned manifold. Consequently, the output provides not just a set of discrete flux values, but a continuous physical model of the surface, enabling the direct measurement of diagnostic parameters such as spectral slope gradients and absorption features that are otherwise inaccessible from photometry alone.

\textcolor{black}{To rigorously quantify the uncertainties associated with this continuous model, we generate an ensemble of reconstructions by sampling the joint posterior distribution of the latent coefficients. We implement a Gaussian Copula framework for this sampling process. Although the PCA components are mathematically orthogonal by construction, the physical distribution of the TNO population in the latent space often retains complex dependencies. By coupling the individual marginal posterior distributions with a structured correlation matrix, the Gaussian Copula explicitly preserves these intrinsic relationships. This mathematically guarantees that our posterior sampling remains strictly within the physically viable regions of the learned manifold, preventing the generation of unphysical spectral realizations while providing robust, statistically sound confidence intervals for the reconstructed features.}

\subsection{Taxonomy Classifier}
In addition to spectral reconstruction, our pipeline incorporates a probabilistic classifier that operates directly within the latent manifold. Rather than attempting to map sparse photometry to discrete taxonomic labels, we perform classification on the predicted PCA mean coefficients. 
This latent space classification applies the dimensionality reduction step as a feature extraction filter: because the principal components encode physical variances such as slope and the molecules spectral features, the decision boundaries between taxonomic groups become significantly more distinct in this coordinate system than in raw color-color space. We trained a secondary AutoML classifier to partition this manifold based on the labels of the training population. Therefore, for every new observation, the pipeline provides not only a reconstructed spectrum but also a probability distribution over spectral types, allowing for the confident identification of classifications.

\section{Training, Validation, and Model Calibration} \label{sec:perf}

This section rigorously tests whether sparse broadband photometry can reliably constrain the low-dimensional spectral manifold derived from near-infrared observations. We evaluate dimensional sufficiency, regression stability, posterior calibration, and classification separability under realistic noise conditions. These diagnostics collectively determine whether the learned latent representation captures physically meaningful spectral variability rather than statistical artifacts.

\subsection{Latent Spectral Representation}
\begin{figure}
\includegraphics[width=.7\columnwidth]{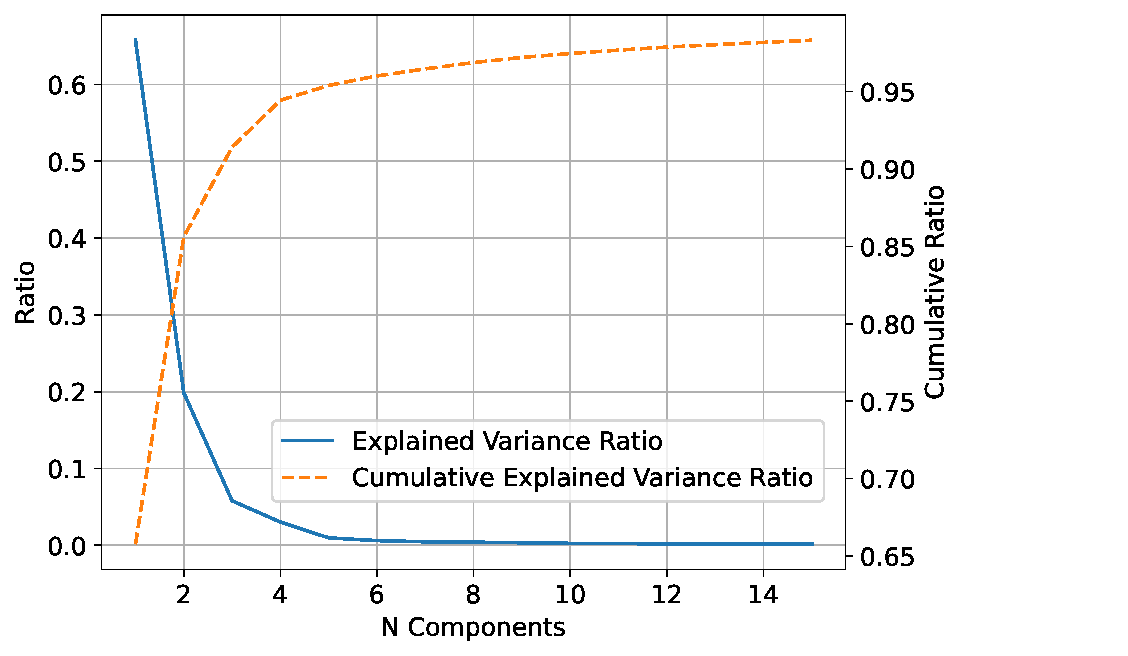}
\centering
\caption{Fractional and cumulative explained variance of the observational seed spectra as a function of the number of principal components. The rapid decay of individual variance (solid blue line) demonstrates that the dominant modes of TNO spectral diversity are intrinsically low-dimensional, with the first five components capturing the vast majority of the variance.}
\label{fig:exp_var}
\end{figure}

We formulated spectral inference in general form as equation~\ref{eq:bayesian_spectral} in Section~\ref{sec:method}, where $S(\lambda)$ denotes the unknown reflectance spectrum and $\mathbf{m}$ the observed photometry. In practice, we restrict $S$ to a low-dimensional spectral manifold learned from the training sample described in Section~\ref{sec:prior}. PCA is applied to the mean-centered spectral matrix to obtain an orthonormal basis $\{\mathbf{e}_k(\lambda)\}$ ordered by decreasing explained variance.

\begin{equation}
S(\lambda) \equiv \bar{f}(\lambda) + \sum_{k=1}^{K} z_k e_k(\lambda).
\end{equation}
where $\bar{f}(\lambda)$ is the sample mean spectrum and $z_k$ are the latent coefficients. This formulation defines a linear spectral manifold embedded in the full wavelength space, with dimensionality $K$.
Therefore, the inverse problem of predicting the high-resolution spectrum $S$ reduces to inferring the posterior distribution of the latent coefficients $\mathbf{z}$ given the observed photometry $\mathbf{m}$.
The leading components capture the large-scale spectral slope and broad absorption morphology, while higher-order components encode progressively finer structural variation. As shown in Figure~\ref{fig:exp_var}, the rapid decay of explained variance indicates that the dominant modes of spectral diversity are intrinsically low-dimensional, motivating a truncated representation.

To evaluate the robustness of the latent basis, we monitor the eigenvectors during the cross-validation process described in Section~\ref{sec:LOOCV}. The first several components remain highly consistent across realizations, with minimal variation in loading structure, while higher-order components exhibit the expected sensitivity to sample perturbations. The appropriate dimensionality $K$ is determined in Section~\ref{sec:k_selection} based on predictive calibration and reconstruction performance.

\subsection{Leave-One-Out Reconstruction Framework}
\label{sec:LOOCV}
To evaluate predictive performance and strictly eliminate data leakage, we adopt a rigorous leave-one-out cross-validation (LOOCV) strategy. Validating a generative model trained on a small, augmented dataset requires ensuring that the model does not merely memorize the local geometries of the synthetic prior. Therefore, for each of the $N$ objects in our reference sample, the target object is completely quarantined. For each LOOCV iteration, the framework executes the following steps strictly on the remaining $N-1$ spectra:
\begin{itemize}
    \item Prior Definition: PCA is recomputed on the $N-1$ spectra to define the latent basis without influence from the held-out object.
    \item Manifold Augmentation: Kernel Density Estimation (KDE) is applied to the $N-1$ latent coordinates to generate thousands of synthetic spectra, mapping the continuous probability distribution of the truncated manifold.
    \item Likelihood Simulation: Synthetic photometry is generated for the KDE-augmented spectra by convolving them with the target filter throughputs and injecting observational Gaussian noise.
    \item Encoder Training: An Automated Machine Learning (AutoML) regressor ensemble is trained from scratch to map the noisy synthetic photometry ($\mathbf{m}$) to the latent coefficients ($\mathbf{z}$).
\end{itemize}
Once the manifold is defined and the regressor is trained, the sparse photometry of the quarantined target object is passed into the pipeline. 

Rather than assuming a rigid analytical Gaussian posterior, we utilize the AutoML framework to empirically estimate the predictive distribution. The gradient-boosted tree ensembles provide not only the expected mean for each latent coefficient but also multi-quantile predictions to capture the statistical uncertainty of the mapping. These predicted latent coordinates are then projected back through the spectral basis $\{\mathbf{e}_k(\lambda)\}$ to generate a predictive distribution over the full wavelength grid. 

This procedure yields both a point estimate (the posterior mean spectrum) and calibrated empirical credible intervals for each reconstructed spectrum. All performance metrics reported below, including reconstruction error and empirical coverage, are computed strictly using the predictions from this fully quarantined LOOCV framework.

\textcolor{black}{Furthermore, the computational expense of this framework is well-suited for standard research hardware. The tabular AutoML algorithms used here are natively CPU-optimal, avoiding the heavy overhead of deep learning. The primary cost involves the posterior sampling required for uncertainty calibration across all K principal components, which takes approximately K minutes for a complete training and sampling cycle. While the total LOOCV process for this study requires roughly $51 \times \text{K}$ minutes $\sim $K hours on a single machine, this is a one-time cost for method validation. In practical applications, the model needs to be trained only once, after which forward inference for new photometric observations is virtually instantaneous.}

\subsection{Dimensionality Selection via Predictive Calibration}
\label{sec:k_selection}

The number of retained principal components $K$ determines the flexibility of the spectral manifold and directly impacts predictive uncertainty. If $K$ is too small, the latent representation cannot capture intrinsic spectral variability, leading to biased reconstructions and underestimated uncertainty. Conversely, excessively large $K$ may introduce poorly constrained directions in latent space without improving predictive performance. We therefore determine $K$ using empirical predictive calibration rather than explained variance alone.

For each choice of $K$, we evaluate reconstruction performance under the LOOCV framework described above. Two complementary diagnostics are considered: (1) reconstruction error relative to the held-out spectrum, and (2) empirical coverage of posterior predictive intervals.
Empirical coverage is defined as the fraction of wavelength points for which the true held-out spectrum lies within the nominal 95\% posterior credible interval. A well-calibrated model should achieve coverage consistent with its nominal level within statistical uncertainty.

We find that low-dimensional representations ($K \lesssim 5$) exhibit systematic under-coverage ($\sim83\%$, when $K = 5$), indicating that the posterior variance is underestimated relative to the residual spectral variability. Increasing $K$ continues to improve the predictive calibration. Specifically, for $K = 8$, the empirical coverage approaches $\sim93\%$, and for $K = 10$, the coverage successfully reaches the nominal $\sim 95\%$ level across objects and wavelengths. Beyond this dimensionality, improvements in predictive coverage are marginal relative to the added model complexity.

These results indicate that insufficient latent dimensionality leads to a regime in which posterior variance is small even though residual variance remains large, particularly in regions of strong spectral curvature. Increasing $K$ reduces this mismatch by allowing the manifold to capture higher-order spectral variation.
We therefore adopt $K = 10$ for subsequent analysis, as it provides near-nominal predictive calibration while maintaining a compact and interpretable latent representation.

We further examine the relationship between latent dimensionality and taxonomic discrimination. Projection of the sample into the leading principal component subspace reveals that class separability is largely established within the first $\sim$3--5 components (see Figure~\ref{fig:latent}). Higher-order components primarily encode finer-scale spectral variation and do not introduce substantial additional class separation. Consistent with this geometric structure, classification accuracy exhibits only marginal improvement beyond $K \approx 4$, despite continued gains in reconstruction calibration for larger $K$.

This behavior reflects a distinction between discriminative and generative objectives: taxonomic classification depends primarily on between-class variance captured in the dominant components, whereas accurate spectral reconstruction and calibrated uncertainty require modeling within-class variability, which resides in higher-order modes. The adopted dimensionality therefore prioritizes predictive calibration over marginal gains in classification accuracy. \textcolor{black}{In physical terms, components $K \le 4$ establish the macro-structure of the spectrum (e.g., overall continuum slopes and broad absorption bowls), while components $K > 4$ map the micro-structure (e.g., fine molecular absorption details). However, because these higher-order modes are more susceptible to non-linear mixing effects and observational noise, inferring precise molecular abundances solely from sparse photometry is intrinsically risky, reinforcing the need for a truncated, regularized manifold.}

\subsection{Wavelength-Dependent Predictive Calibration}
\label{sec:wcal}
While global coverage provides an overall assessment of calibration, spectral reconstruction quality varies systematically with wavelength. To quantify this behavior, we compute empirical coverage as a function of wavelength across the LOOCV sample.
At each wavelength $\lambda_j$, coverage is defined as the fraction of held-out spectra for which the true reflectance lies within the nominal 95\% posterior credible interval. A well-calibrated model should exhibit coverage consistent with its nominal level within binomial uncertainty.

We find that calibration is wavelength-dependent. Regions characterized by smooth continuum behavior exhibit near-nominal coverage, while under-coverage is more pronounced in high-curvature regions associated with strong absorption features. This effect is most significant for lower-dimensional latent representations and diminishes as $K$ increases.
This behavior reflects a regime in which posterior variance may be small even though residual variance remains large, indicating that the linear manifold insufficiently captures localized spectral nonlinearity. Increasing $K$ allows higher-order structure to be represented explicitly, reducing the discrepancy between posterior and residual variance.

\begin{figure}
\includegraphics[width=.7\columnwidth]{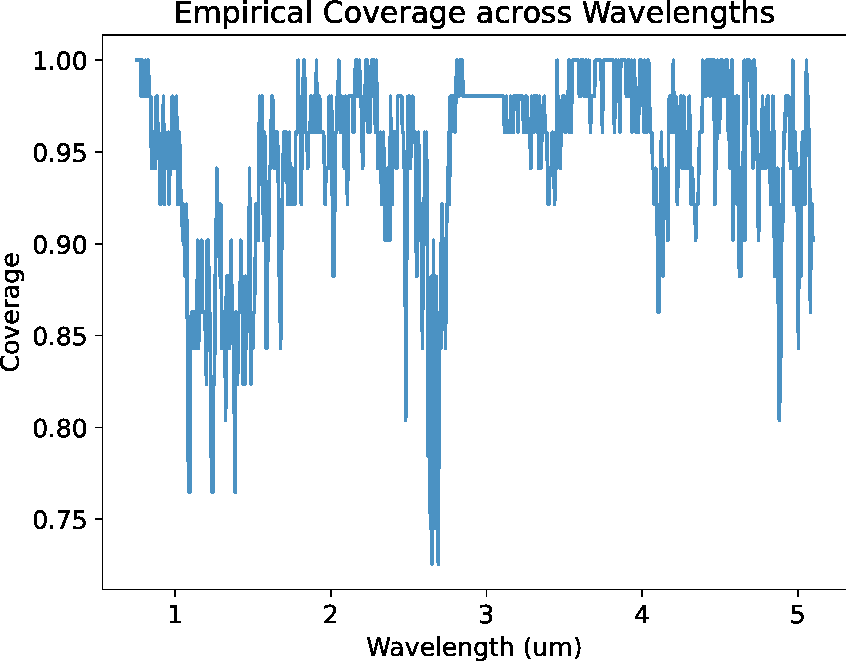}
\centering
\caption{Empirical coverage as a function of wavelength for nominal 95\% posterior credible intervals under LOOCV. Coverage is computed as the fraction of held-out spectra whose true reflectance lies within the reconstructed interval at each wavelength. Coverage remains near nominal across most wavelengths, with localized decreases near the strong absorption features, particularly around 3$\mu$m where no direct photometric sampling is available, reflecting reduced likelihood constraints in sparsely sampled regions.}
\label{fig:coverage}
\end{figure}

Figure~\ref{fig:coverage} shows wavelength-dependent empirical coverage for nominal 95\% posterior credible intervals. For most wavelengths, coverage remains between 90\% and 100\%, consistent with nominal calibration within binomial uncertainty for a sample of 51 objects. The localized decrease in empirical coverage near 2.7–2.9 $\mu$m is expected, as this region is not directly sampled by the adopted photometric filters\footnote{All reconstructions evaluated here adopt the four-band configuration [F090W, F115W, F410M, F460M].}. In the absence of direct observational constraints, the posterior distribution in this wavelength range is governed primarily by the prior covariance structure of the latent manifold. If the held-out object exhibits stronger-than-average curvature in this region, the truncated representation may underestimate the true variance, leading to modest under-coverage. This behavior reflects the intrinsic limitations of sparse photometric sampling and truncated latent dimensionality, rather than a breakdown of the Bayesian formulation.

\subsection{Reconstruction Accuracy}

In addition to empirical coverage, we evaluate reconstruction fidelity using the root-mean-square error (RMSE) between the posterior mean spectrum and the held-out ground-truth spectrum under LOOCV.

Reconstruction error decreases monotonically with increasing $K$, reflecting improved representation of higher-order spectral variation. However, the reduction in RMSE beyond $K \approx 6$ becomes effectively flat. This behavior mirrors the calibration analysis of Section~\ref{sec:k_selection} and supports the adopted dimensionality.
Importantly, improvements in RMSE are accompanied by improved predictive calibration, indicating that increased model flexibility enhances both accuracy and uncertainty reliability rather than merely overfitting the data.

\begin{figure}
\includegraphics[width=.7\columnwidth]{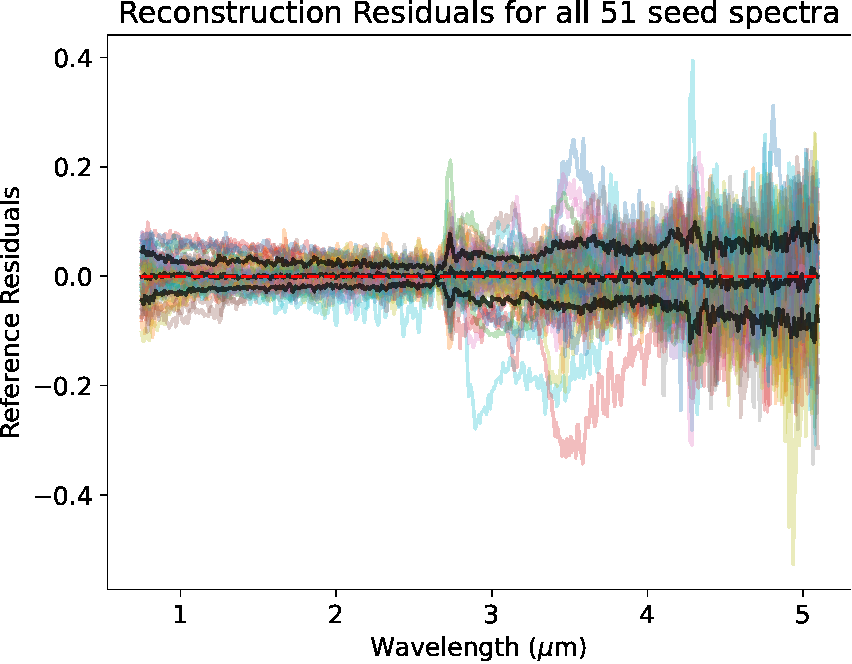}
\centering
\caption{Figure 5: Leave-one-out cross-validation (LOOCV) reconstruction residuals (reconstructed minus reference reflectance) for the 51 TNO seed spectra. \textcolor{black}{The black lines indicates the residuals of 16th, 50th, and 84th percentile range (corresponding to a $\pm 1\sigma$ confidence interval). The residuals are consistently centered near zero (red dash line) across the $0.7$--$5.0~\mu\text{m}$ range, indicating that the reconstruction pipeline is free of significant systematic bias. The increased dispersion observed beyond $3~\mu\text{m}$ is driven by the presence of complex molecular absorption features and higher spectral curvature in certain taxonomic groups, which naturally introduce greater reconstruction variance in the long-wavelength regime.}}
\label{fig:resid}
\end{figure}

Figure~\ref{fig:resid} shows the wavelength-dependent residuals under LOOCV for all seed spectra. Residuals remain centered around zero across the full wavelength range, indicating no significant systematic bias. The dispersion increases toward longer wavelengths, particularly beyond 3$\mu$m, where strong molecular absorption features introduce higher spectral curvature. This behavior is consistent with the wavelength-dependent calibration analysis presented in Section~\ref{sec:wcal} and reflects the greater intrinsic variance of the training sample in these regions.

\begin{figure*}
        \centering
        \begin{subfigure}[b]{0.49\textwidth}
            \centering
            \includegraphics[width=\textwidth]{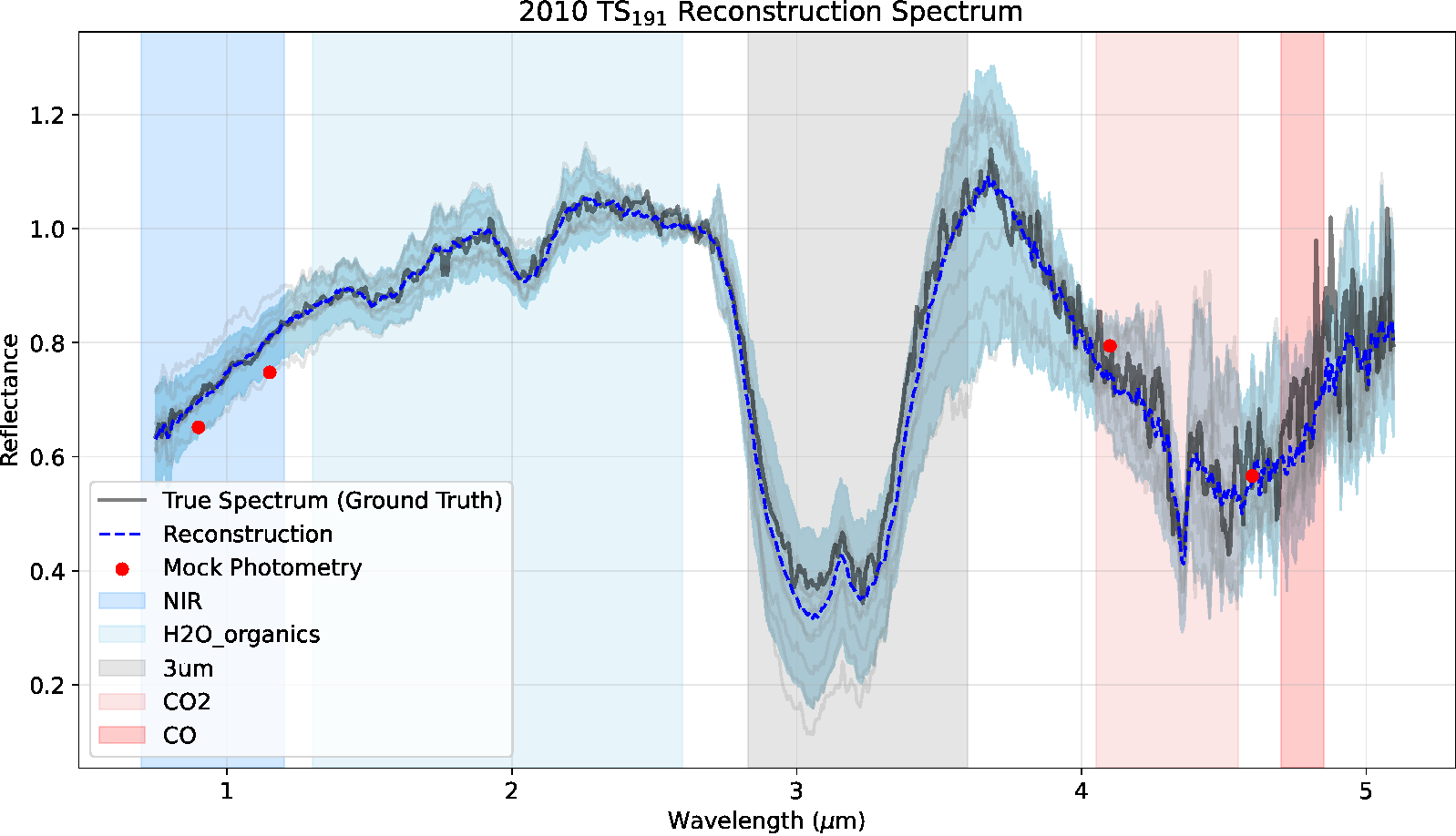}
        \end{subfigure}
        \hfill
        \begin{subfigure}[b]{0.49\textwidth}  
            \centering 
            \includegraphics[width=\textwidth]{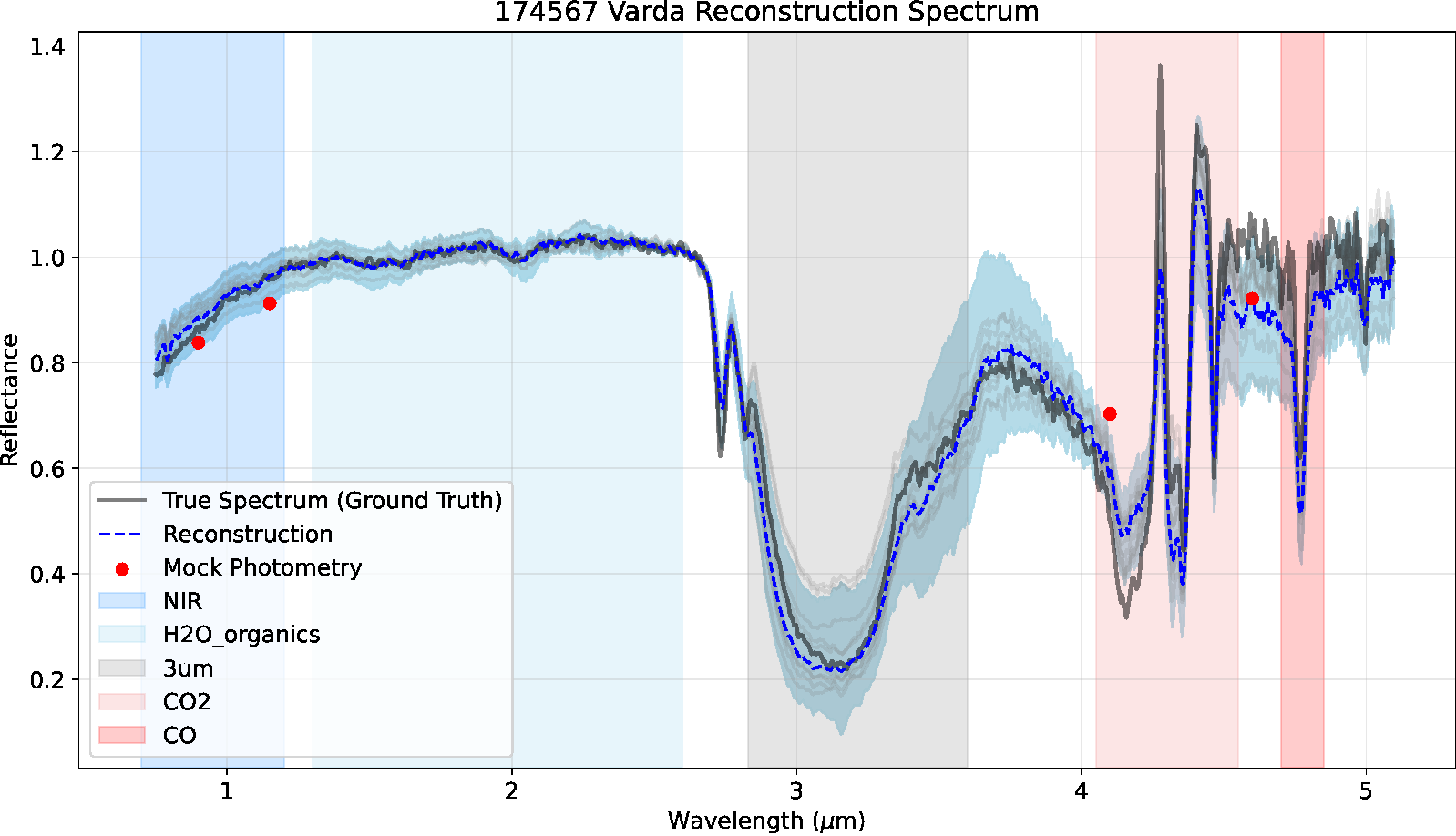}
        \end{subfigure}
        \vskip\baselineskip
        \begin{subfigure}[b]{0.49\textwidth}   
            \centering 
            \includegraphics[width=\textwidth]{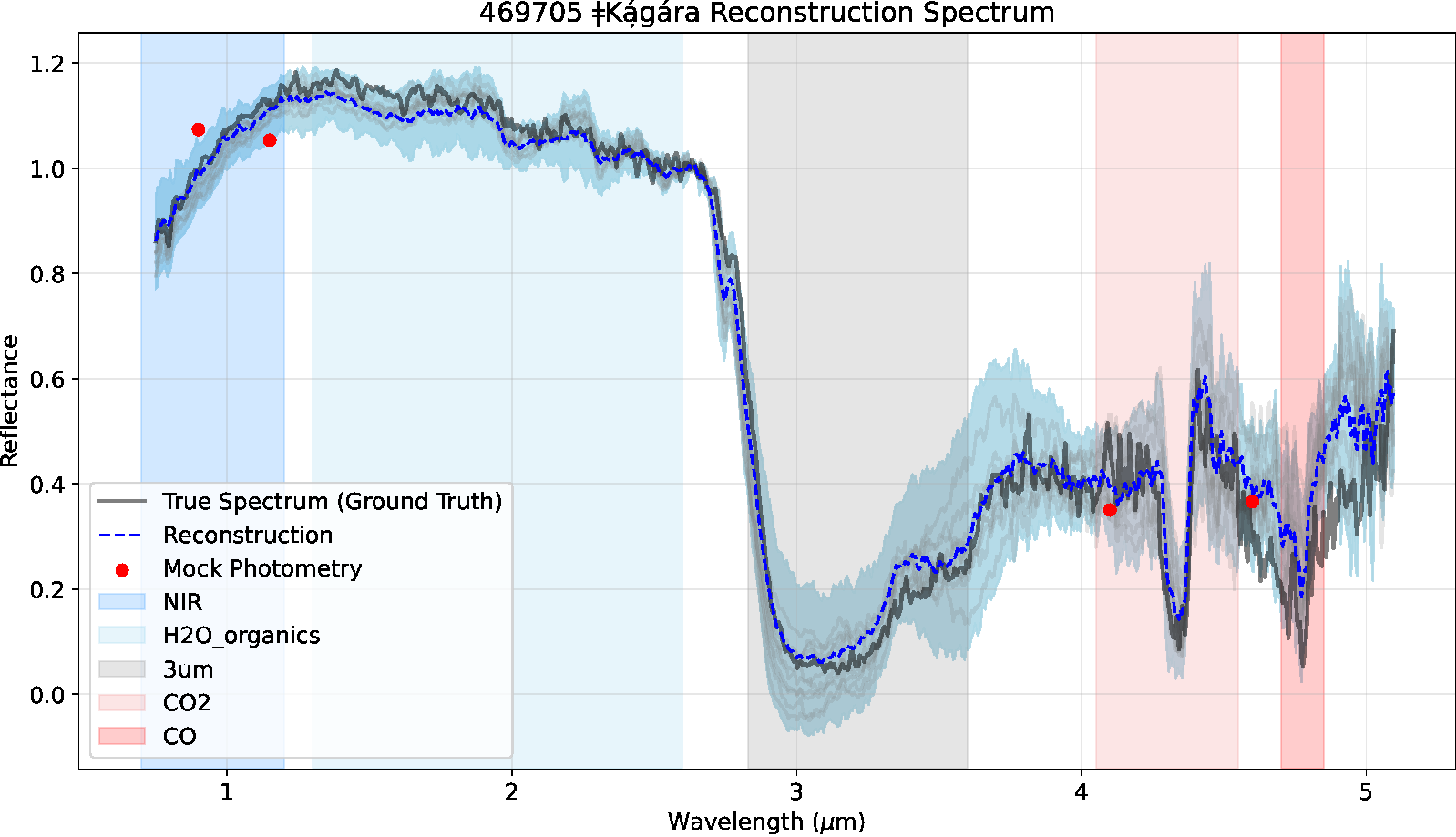}
        \end{subfigure}
        \hfill
        \begin{subfigure}[b]{0.49\textwidth}   
            \centering 
            \includegraphics[width=\textwidth]{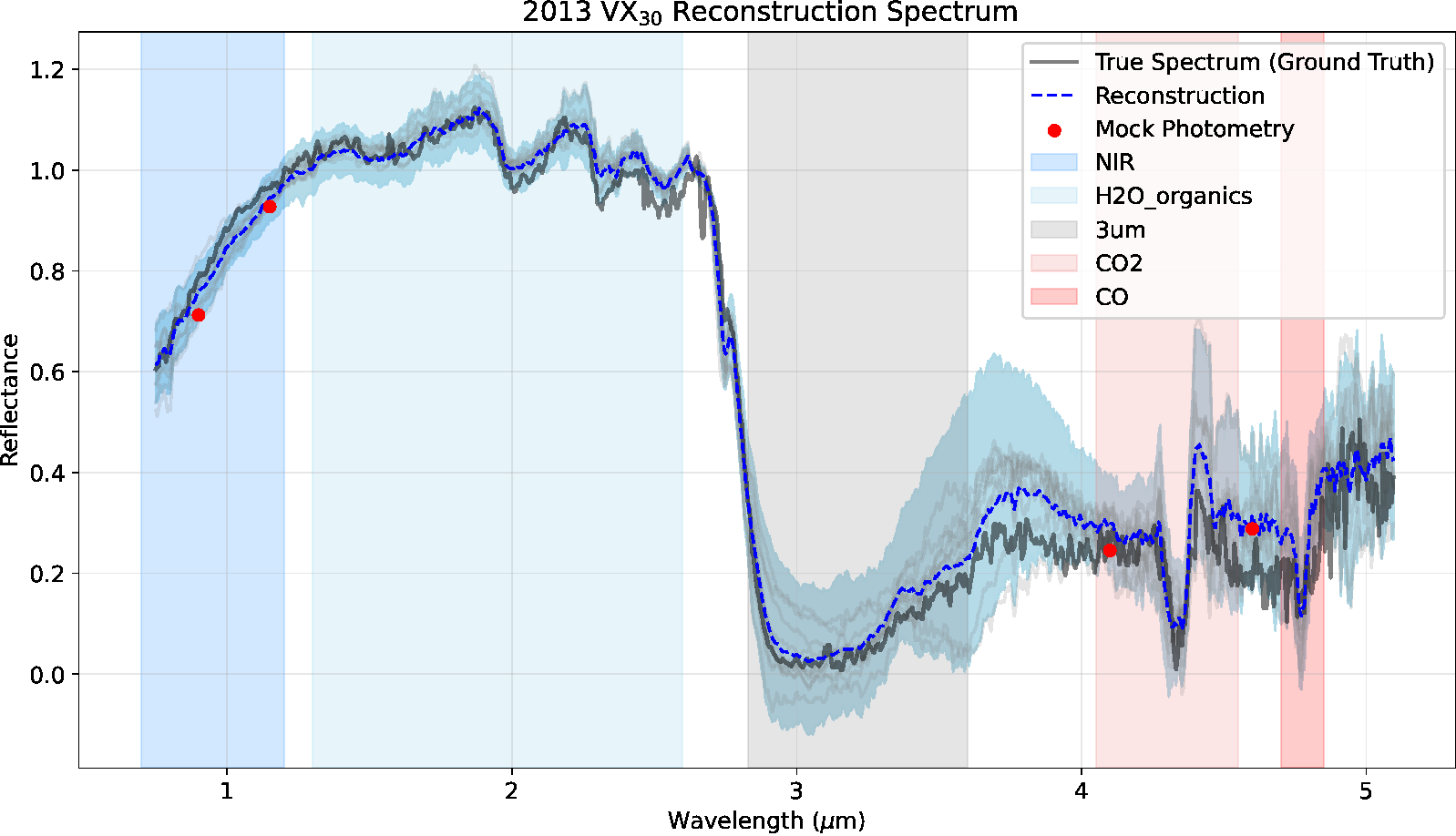}
        \end{subfigure}
        \caption{Reconstructed spectra for four distinct TNO spectral types. The spectra of the two Neptune Trojans, 2010~TS$_{191}$ and 2013~VX$_{30}$, were withheld from the training seeds to test generalization. The two TNOs 174567~Varda and 469705~\textdoublebarpipe K\cb{a}g\cb{a}ra were reconstructed using a Leave-One-Out Cross-Validation (LOOCV) framework. The shaded regions indicate the 95\% posterior credible intervals, while the gray lines represent 10 alternative realizations randomly sampled from the posterior. Major absorption features (3 $\mu$m, $\text{H}_2\text{O}$, $\text{CO}_2$, CO) are highlighted.}
        \label{fig:spec}
\end{figure*}

Figure \ref{fig:spec} demonstrates the capability of probabilistic spectral reconstruction from photometry within this framework. The ``Reconstruction'' corresponds to the posterior mean in latent space, projected into wavelength space. This reconstruction remains largely centered within the 95\% posterior credible intervals. Ten alternative realizations, randomly drawn from the posterior manifold, illustrate the range of statistical uncertainty. Key diagnostic bands are highlighted; several major absorption features, including the 3 $\mu$m band, water, CO$_2$, and CO, are probabilistically constrained, with uncertainties naturally increasing in regions of high spectral curvature.

\subsection{Taxonomic Classification Performance}
\label{sec:class_performance}

We evaluate taxonomic classification under the same LOOCV framework used for spectral reconstruction. In each iteration, the held-out object is excluded from the PCA construction, KDE density estimation, synthetic augmentation, regressor training, and classifier training. Classification is then performed on the predicted latent coordinates of the held-out object, ensuring that reported performance reflects true predictive generalization.

To mitigate variance associated with the limited number of seed spectra, we augment the training set within each LOOCV fold by sampling from the KDE prior in latent space and injecting Gaussian photometric noise consistent with the adopted signal-to-noise assumptions. This Monte Carlo augmentation regularizes the regression and classification models by densely sampling the locally supported manifold without introducing information from the held-out object. Across multiple random seeds, classification accuracy varies by less than $\sim$2\%, indicating stability of the learned decision boundaries.

Under the default four-filter configuration adopted throughout Section~\ref{sec:perf} ([F090W, F115W, F410M, F460M]), LOOCV classification accuracy reaches $\sim95\%$ for water-dominated and CO$_2$-dominated objects, $\sim90\%$ for methanol-rich organics, and $\sim85\%$ for methanol-poor organics. Misclassifications occur almost exclusively between the two organic subclasses, with negligible confusion between ice-dominated and organic-dominated classes.

This behavior is directly reflected in the latent-space geometry shown in Figure~\ref{fig:latent}. In the projection onto the leading principal components (PC1, PC2), H$_2$O- and CO$_2$-dominated spectra form compact and well-separated clusters, explaining their high classification accuracy. In contrast, methanol-rich and methanol-poor organic spectra occupy partially overlapping regions of the manifold, particularly along higher-order components (PC3, PC4) that encode subtler variations in continuum slope and absorption curvature. The observed confusion between these subclasses therefore follows naturally from the intrinsic spectral geometry rather than limitations of the classifier architecture.

Consistent with the dimensionality analysis of Section~\ref{sec:k_selection}, classification performance saturates for $K \gtrsim 4$, even though reconstruction calibration continues to improve for larger $K$. This distinction highlights the difference between discriminative and generative objectives: between-class variance is largely captured in the dominant principal components, whereas within-class variability, important for calibrated spectral reconstruction, resides in higher-order modes. A detailed confusion analysis and filter-dependence study are presented in Section~\ref{sec:bands}.

Taken together, these results demonstrate that the near-IR spectral diversity of the current TNO sample is effectively captured by a low-dimensional latent representation that remains statistically well constrained by sparse photometric measurements. The reconstruction is robust under LOOCV, the posterior uncertainties are well calibrated, and classification performance reflects the intrinsic geometric separation of surface types in latent space. The dominant limitations are not from model instability but from filter coverage and the finite diversity of the training manifold. Having established statistical validity, we now explore practical applications of the framework.

\section{Applications} \label{sec:app}

The primary application of this framework is to translate sparse near-IR photometry into probabilistic constraints on TNO taxonomy and spectral morphology, thereby enabling more efficient use of spectroscopic resources. Furthermore, these probabilistic reconstructions have direct utility for future observation planning. Observers can forward-propagate anticipated observational errors by sampling photometric inputs according to their expected signal-to-noise ratios. For each sampled photometric realization, the pipeline generates a corresponding reconstructed spectrum and credible interval. By marginalizing over all realizations for a given simulated observation, one obtains a fully propagated predictive distribution, yielding the final reconstructed spectrum and realistic uncertainty bounds tailored to the planned observation.

A critical question therefore arises for future survey design: which filters yield the highest information content? In other words, what are the minimal optimal filter sets required to distinguish TNOs into their major spectral types?

\subsection{Survey optimization analysis: The Most Informative Bands} \label{sec:bands}

The unique design of JWST/NIRCam, which utilizes a dichroic to split light into short-wavelength (SW) and long-wavelength (LW) channels, enables the simultaneous acquisition of two photometric bands. Here, we test configurations corresponding to one, two, and three exposures (yielding 2, 4, and 6 color-indices, respectively) to determine the optimal filter sets. We evaluated all possible combinations of wide (W) and medium (M) filters, excluding narrow bands and the F070W filter (centering at 0.7 $\mu$m) because our training spectra do not uniformly cover the full F070W bandpass. Detailed instrument specifications are drawn from the JWST/NIRCam documentation\footnote{\url{https://jwst-docs.stsci.edu/jwst-near-infrared-camera}}.

\subsubsection{Two-Filter Combinations (Single Exposure)}
\begin{figure}
\includegraphics[width=.24\columnwidth]{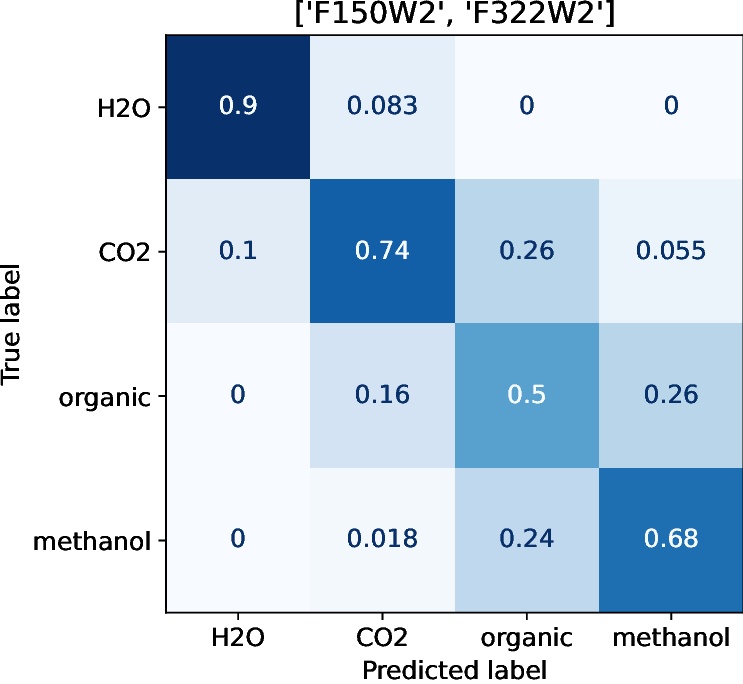}
\includegraphics[width=.24\columnwidth]{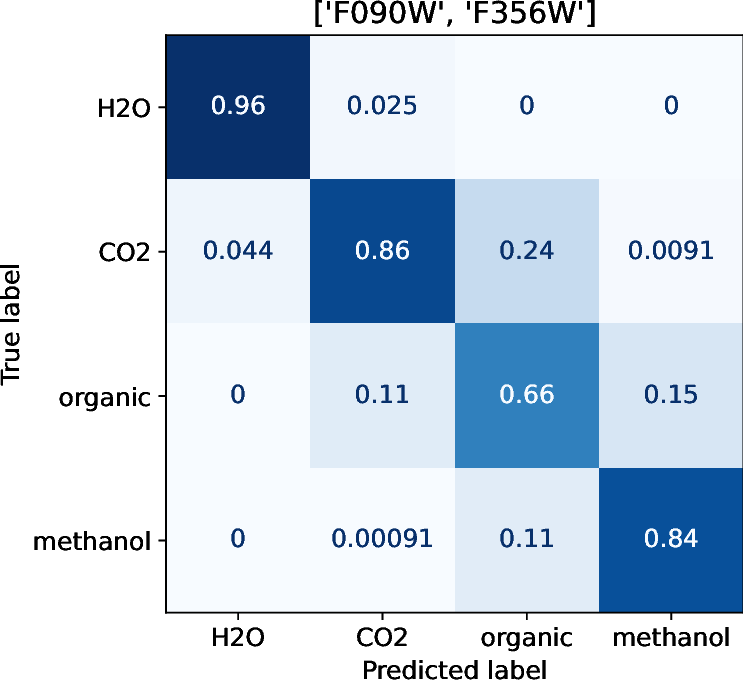}
\includegraphics[width=.24\columnwidth]{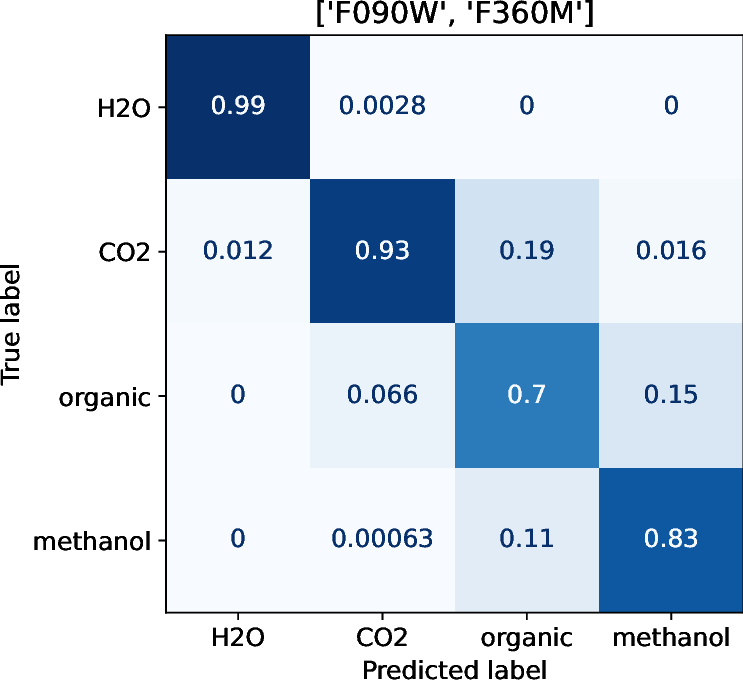}
\includegraphics[width=.24\columnwidth]{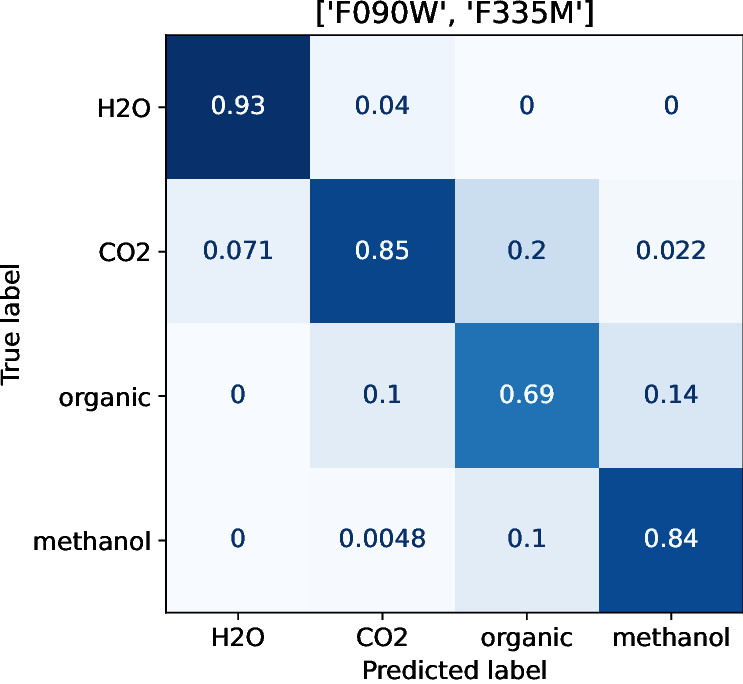}
\caption{Confusion matrices for four selected 2-band photometry classification scenarios.}
\label{fig:2bands}
\end{figure}

With a single exposure obtaining one SW and one LW band, we find that even with minimal resources, an optimal filter choice yields fair classification accuracy. Figure~\ref{fig:2bands} displays the confusion matrices for selected two-filter combinations. By utilizing the two broadest bands available, F150W2 and F322W2, water-type spectra are identified with high accuracy (Figure~\ref{fig:2bands}, leftmost panel). The CO$_2$ and Organic types are also identified with moderate success (70--80\% accuracy); however, methanol-poor organic types are frequently confused with CO$_2$ and methanol-rich organic types. We note that the [F150W2, F322W2] combination was employed in JWST GO programs Cycle 1 \#1598 \citep{jwst1568} and Cycle 4 \#7700 \citep{jwst7700}. While these programs were designed for ultra-deep discovery surveys, our results suggest they can serendipitously perform robust classification for newly discovered TNOs, provided these small objects stick to the three major spectral types identified by \citet{disco2024}.

When utilizing the standard wide (W) filters, we find that the most informative SW filter is F090W (centered at 0.9 $\mu$m), followed by F115W (centered at 1.15 $\mu$m). Coupling these LW filters with a 3.5 $\mu$m filter, such as F356W (Figure~\ref{fig:2bands}, second panel from left), creates a robust configuration. The ability to identify CO$_2$ types and separate the two organic subtypes is significantly improved compared to the ultra-wide [F150W2, F322W2] combination. Swapping F356W for the narrower F360M further improves accuracy (Figure~\ref{fig:2bands}, third panel). However, replacing F360M with F335M degrades performance back to the level of the F090W/F356W combination (Figure~\ref{fig:2bands}, rightmost panel). Based on these tests, we identify the following two-band configurations as optimal within the current latent framework [F150W2, F322W2] (for maximum photon efficiency), [F090W, F356W] (for a balance of depth and accuracy), and [F090W, F360M] (for maximum classification accuracy).

\begin{figure}
\includegraphics[width=.7\columnwidth]{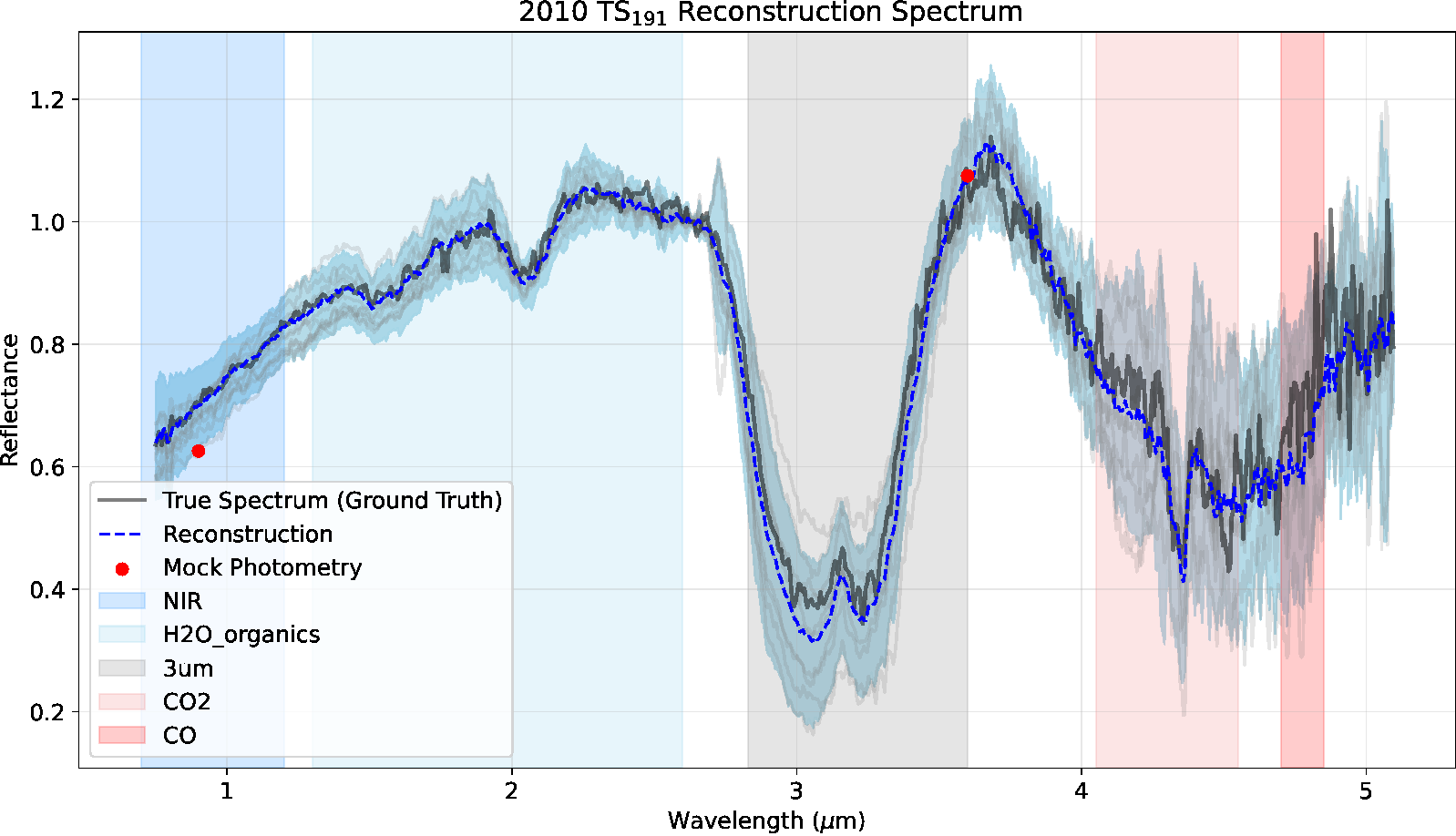}
\centering
\caption{Example of a water-type spectrum reconstruction using only [F090W, F360M] photometry. }
\label{fig:spec_2bands}
\end{figure}

Regarding spectral reconstruction with two filters, while the broad continuum shape is generally recovered correctly, fine-scale details are often have larger uncertainty due to the weak constraints provided by only two data points. Figure \ref{fig:spec_2bands} illustrates a reconstruction using the F090W/F360M combination. Although the model successfully identifies the object as a water-type, the specific profiles of diagnostic absorption features (such as the 3.0 $\mu$m absorption band and 4.26 $\mu$m CO$_2$ band) remain inaccurate. However, we note a distinct advantage at the specific wavelengths sampled by the photometry: because the F360M filter directly anchors the model near 3.6 $\mu$m, the predictive uncertainty ``pinches'' and becomes narrow at this location. Consequently, a targeted 2-band reconstruction can actually outperform a 4-band configuration at specific wavelengths if the latter omits a local filter, as the direct photometric constraint successfully collapses the local posterior variance. Thus, while two-band reconstruction lacks the global fidelity for comprehensive feature characterization, it provides highly precise, localized anchors for the spectral continuum.

\subsubsection{Four-Filter Combinations (Two Exposures)}
While four-filter photometry is generally more effective than two-filter sets, performance remains highly dependent on filter selection. We find that the hierarchy of information content for SW filters is F090W $>$ F115W $>$ F150W. Other filters, including the narrower M bands covering the 1.5--2.0 $\mu$m range (typically used for water and organic absorption diagnosis), are surprisingly less effective for global classification than their shorter wavelength counterparts in this sparse framework. For the LW channel, the most informative filter is F460M, followed by F410M, and then F360M/F356W.

\begin{figure}
\includegraphics[width=.24\columnwidth]{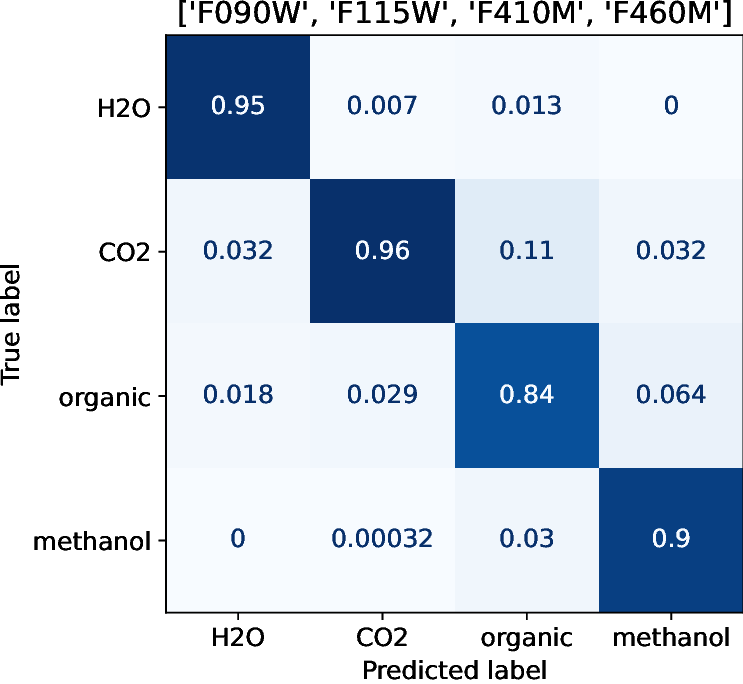}
\includegraphics[width=.24\columnwidth]{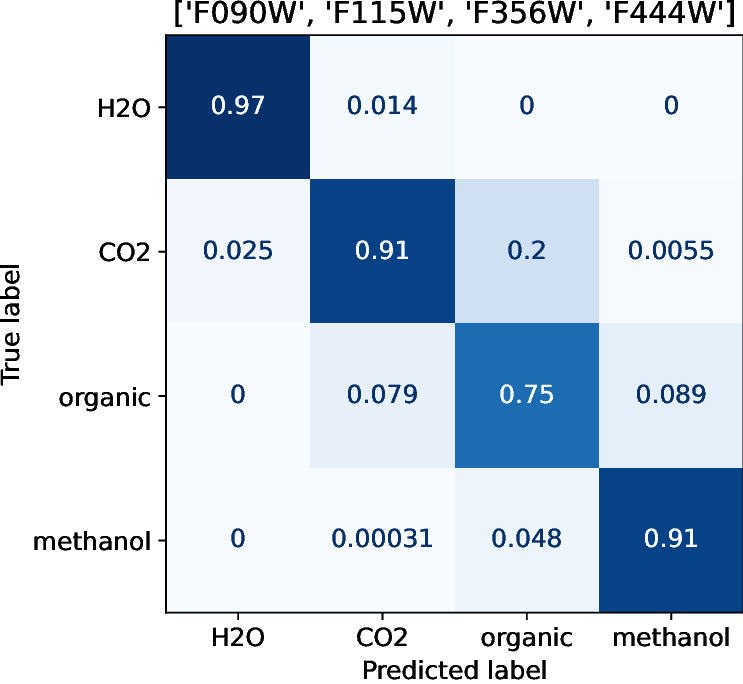}
\includegraphics[width=.24\columnwidth]{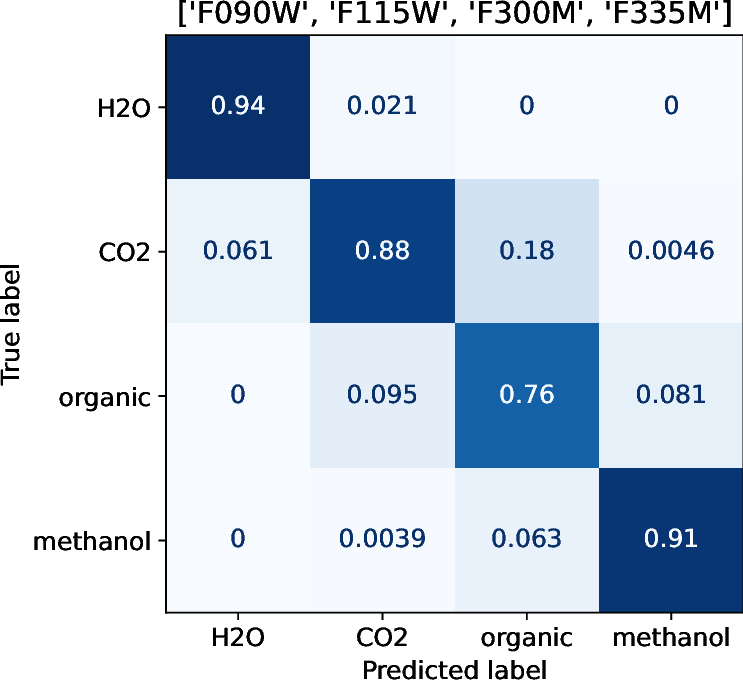}
\includegraphics[width=.24\columnwidth]{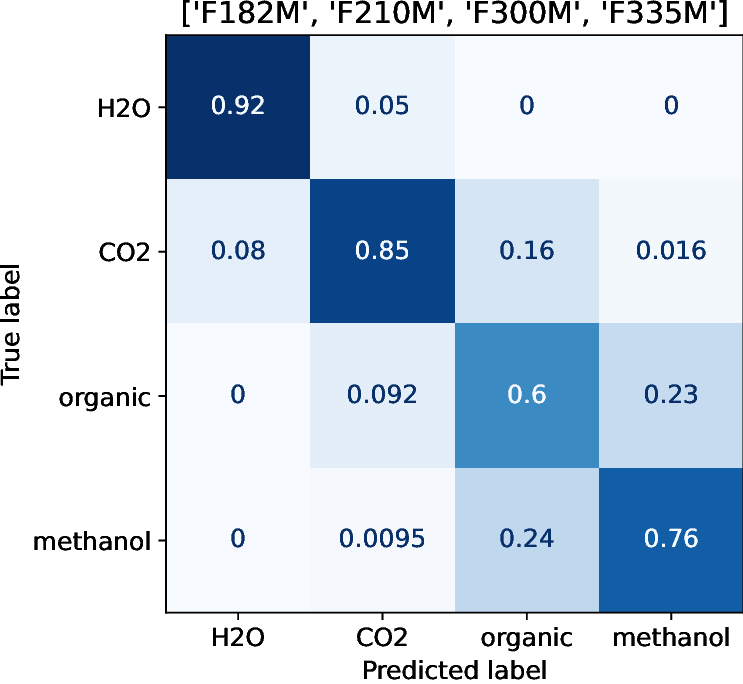}

\caption{Confusion matrices for four selected 4-band photometry classification scenarios. The rightmost panel corresponds to the filter set of JWST GO \#7248.}
\label{fig:4bands}
\end{figure}

The combination of F090W, F115W, F410M, and F460M represents a powerful photometric set for TNO taxonomy. The leftmost panel of Figure~\ref{fig:4bands} shows the confusion matrix for this set, where accuracy exceeds 90\% across all four spectral types. Alternatively, one can trade maximal accuracy for higher sensitivity by using the broader F356W and F444W filters (Figure~\ref{fig:4bands}, second panel).

There are compelling observational reasons to avoid filters with wavelengths longer than 3.5 $\mu$m: (1) the solar flux decreases significantly, and (2) thermal emission from the TNO itself may contaminate the reflectance signal. Therefore, the combination of [F090W, F115W, F300M, F335M] can be a high-priority consideration. Due to the higher solar flux available when limiting the longest wavelength to $\sim$3.35 $\mu$m, this configuration rivals the efficiency of the [F090W, F115W, F356W, F444W] set while simplifying the physical interpretation without the need for modeling thermal flux (pure reflectance).

Finally, we comment on the filter choice [F182M, F210M, F330M, F335M] employed by JWST Cycle 4 GO program \#7248 \citep{jwst7248}. Our analysis indicates that the most informative SW filters for TNO taxonomy are F090W and F115W; utilizing intermediate bands offers no significant gain in discriminatory power under this latent space framework. We find that while the GO \#7248 filter set is effective for distinguishing Water, CO$_2$, and Organic types, swapping [F182M, F210M] for [F090W, F115W] would significantly improve the efficiency of separating the two subtypes (methanol-rich vs. methanol-poor) of the Organic group (Figure~\ref{fig:4bands}, rightmost panel). Furthermore, our models suggest the four-band [F182M, F210M, F330M, F335M] combination does not perform significantly better on taxonomy than the optimized \textbf{two-band} combinations, such as [F090W, F360M] and [F090W, F335M] suggested above.

\subsubsection{Six-Filter Combinations (Three Exposures)}
\begin{figure}
\centering
\includegraphics[width=.3\columnwidth]{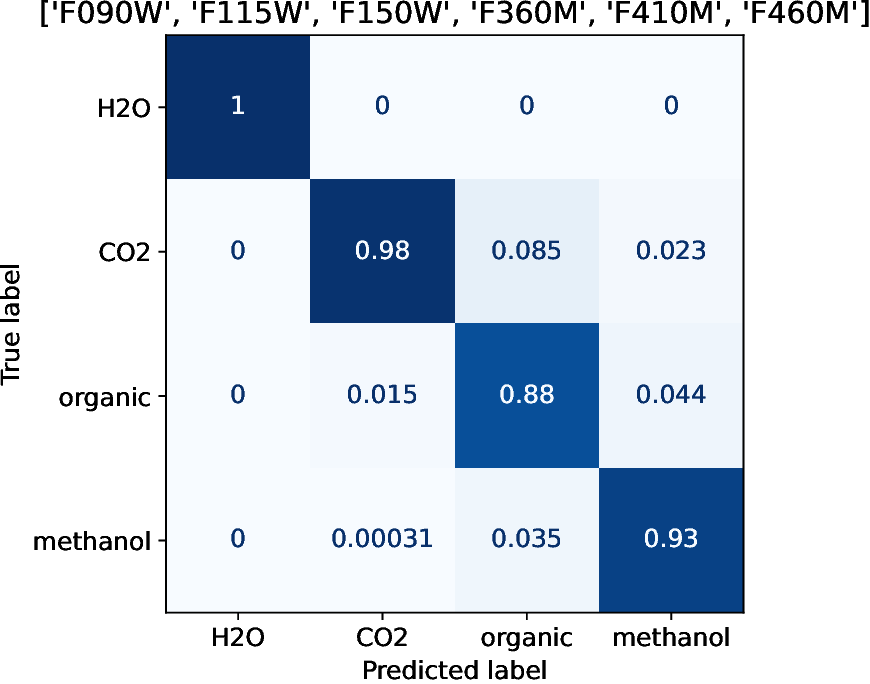}
\includegraphics[width=.3\columnwidth]{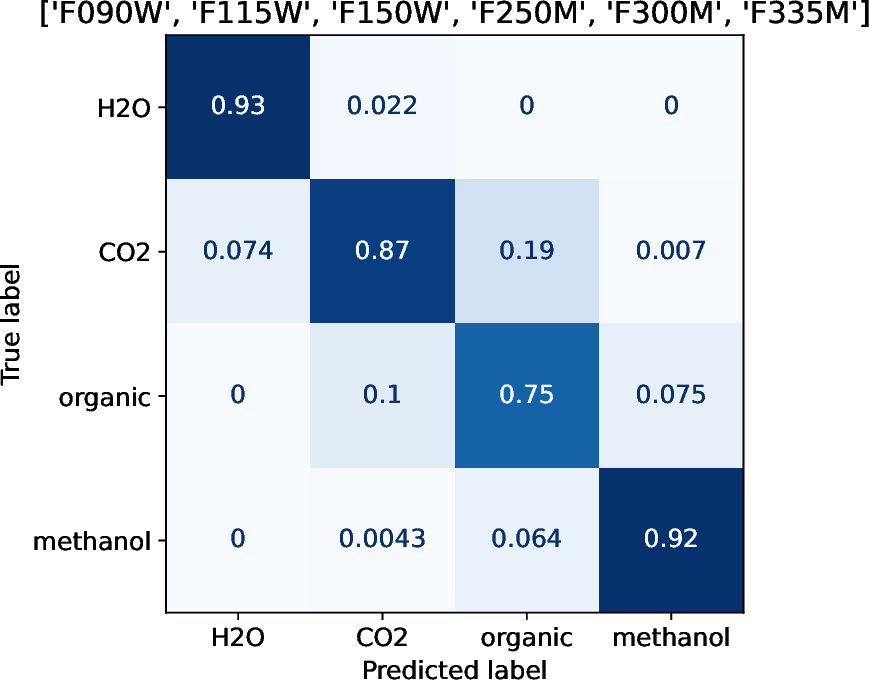}
\includegraphics[width=.3 \columnwidth]{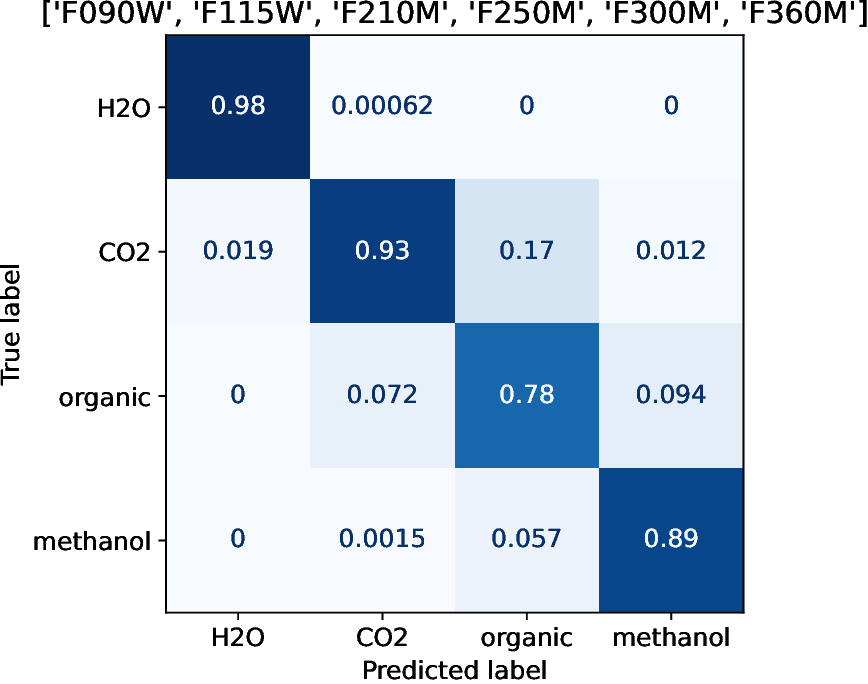}
\caption{Confusion matrices for four selected 6-band photometry classification scenarios.}
\label{fig:6bands}
\end{figure}

Although one might expect that increasing the number of photometric bands to six would substantially refine the TNO taxonomy, we find the marginal gains to be minimal. Figure~\ref{fig:6bands} illustrates the confusion matrices for three representative six-filter configurations. The sets [F090W, F115W, F150W, F360M, F410M, F460M] and [F090W, F150W, F200W, F277W, F300M, F335M] serve as expanded versions of the four-filter sets discussed previously. The [F090W, F115W, F210M, F250M, F300M, F360M] mimics the choice of JWST cycle 3 GO program \#6064 \citep{jwst6064} with the only difference being the substitution of F090W instead of F070W.   

Comparing these to the four-filter baselines, the improvements in classification accuracy are negligible. We conclude that the additional two filters provide little distinct information regarding the major taxonomic groups defined in the current latent space. However, the true benefit of a six-filter setup likely lies not in classification, but in the enhanced ability to constrain the spectral morphology (Section~\ref{sec:outliers}).

We note that the real-world per-class accuracies may not be as high as presented here, as the results carry binomial uncertainty due to the finite sample size. Nevertheless, the relative ranking of filter configurations is highly stable across the LOOCV folds and reflects the underlying geometric separability of the classes within the latent space rather than stochastic variation. Because the differences in overall classification accuracy between the optimal and sub-optimal filter configurations are typically 5--10\%, exceeding the binomial uncertainty for the dominant compositional classes (see Section~\ref{sec:limitations}), the qualitative hierarchy of these filter sets remains robust.

\subsection{Detecting Rare Spectral Types}
\label{sec:outliers}
Since rare spectral morphotypes were intentionally excluded from the training seed population, ideally, these outliers are absent from the model's prior knowledge. When the pipeline processes the photometry of a spectral outlier, the generative model projects the input onto the nearest point within the learned latent manifold (the prior). This results in a reconstructed spectrum that mimics the outlier's colors using only standard physical components, a process mathematically analogous to \textbf{``deepfakes''} or AI hallucinations, where a generative model forces an out-of-distribution input to conform to a learned prior.

In reality, the situation is complicated. While some outliers, such as Haumea family members (pure crystalline water ice), have spectra distinct from the major groups, others, such as some of the Neptune Trojans, share broad similarities with standard TNOs. During the creation of the synthetic training set via Kernel Density Estimation (see Section~\ref{sec:prior}), the data augmentation process inevitably creates intermediate \textbf{``Frankenstein''} spectra, statistical combinations of the seed types that fill the interstitial latent space. Although these ``Frankenstein'' spectra represent low-probability intermediate regions of the prior, sufficiently constraining photometry can effectively ``awaken'' these intermediate morphologies, causing them to manifest in the posterior reconstruction. This capability suggests the pipeline could recover transitional spectral shapes that do not strictly exist in the discrete seed sample, but lie within the continuous topology of the learned manifold.

To empirically validate this capability, we apply the reconstruction pipeline to two known spectral outliers within the Neptune Trojan population: 2006~RJ$_{103}$ and 2011~SO$_{277}$ (\citealp{Markwardt2025}, \dataset[DOI: 10.17909/j66s-pv96]{https://doi.org/10.17909/j66s-pv96}). The object 2006~RJ$_{103}$ exhibits a spectral morphology resembling water-type TNOs but notably lacks significant water absorption features. The other Neptune Trojan, 2011~SO$_{277}$ displays spectral characteristics intermediate between the major TNO groups, sharing morphological similarities with the ``blue binary'' population of the Cold Classical Kuiper Belt \citep{Wong2025PSJ}. Identifying such minority spectral classes is critical for mapping the dynamical transport of chemically distinct bodies across the Trans-Neptunian region.

\begin{figure}
\centering
\includegraphics[width=.7\columnwidth]{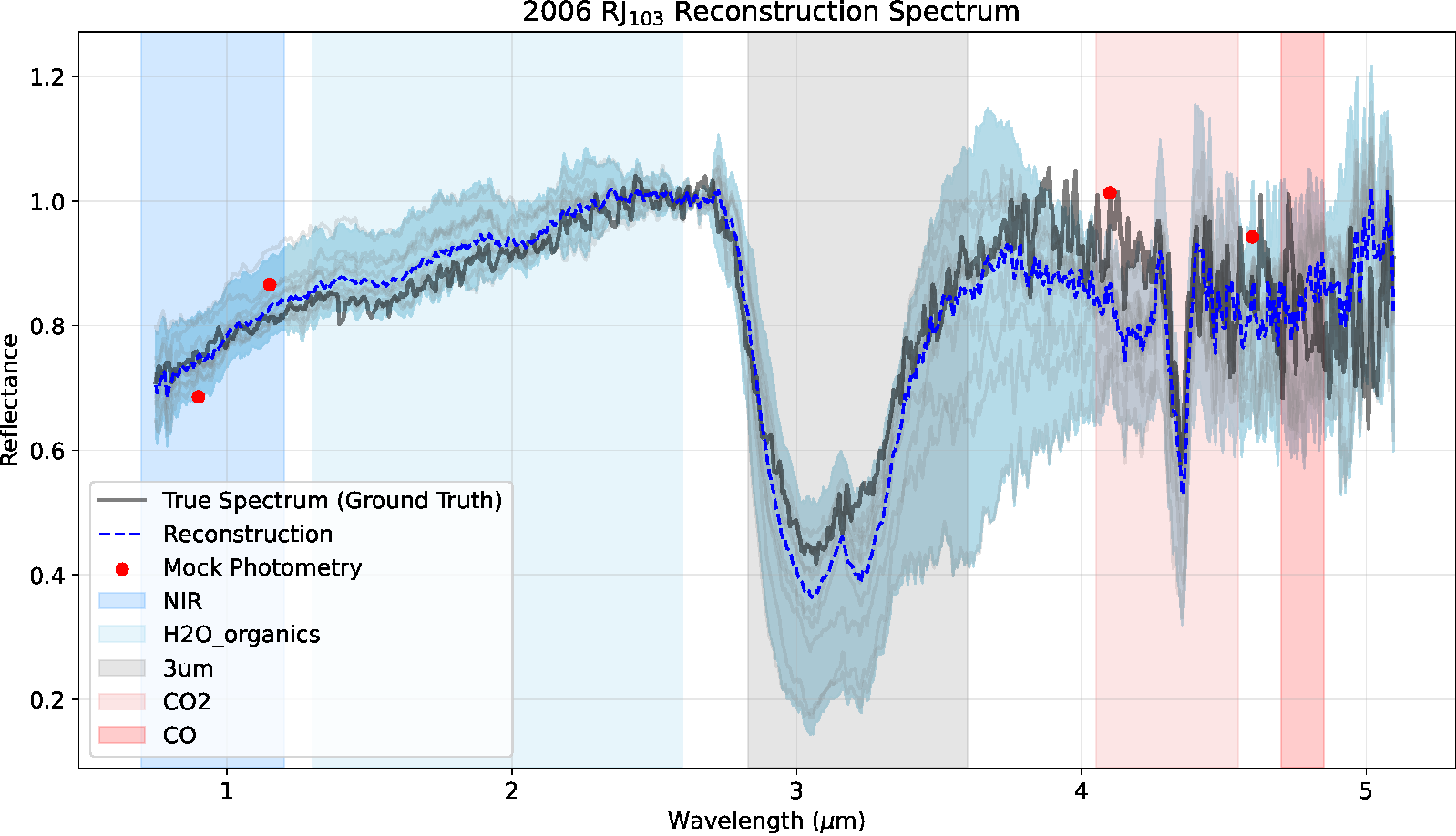}
\includegraphics[width=.7\columnwidth]{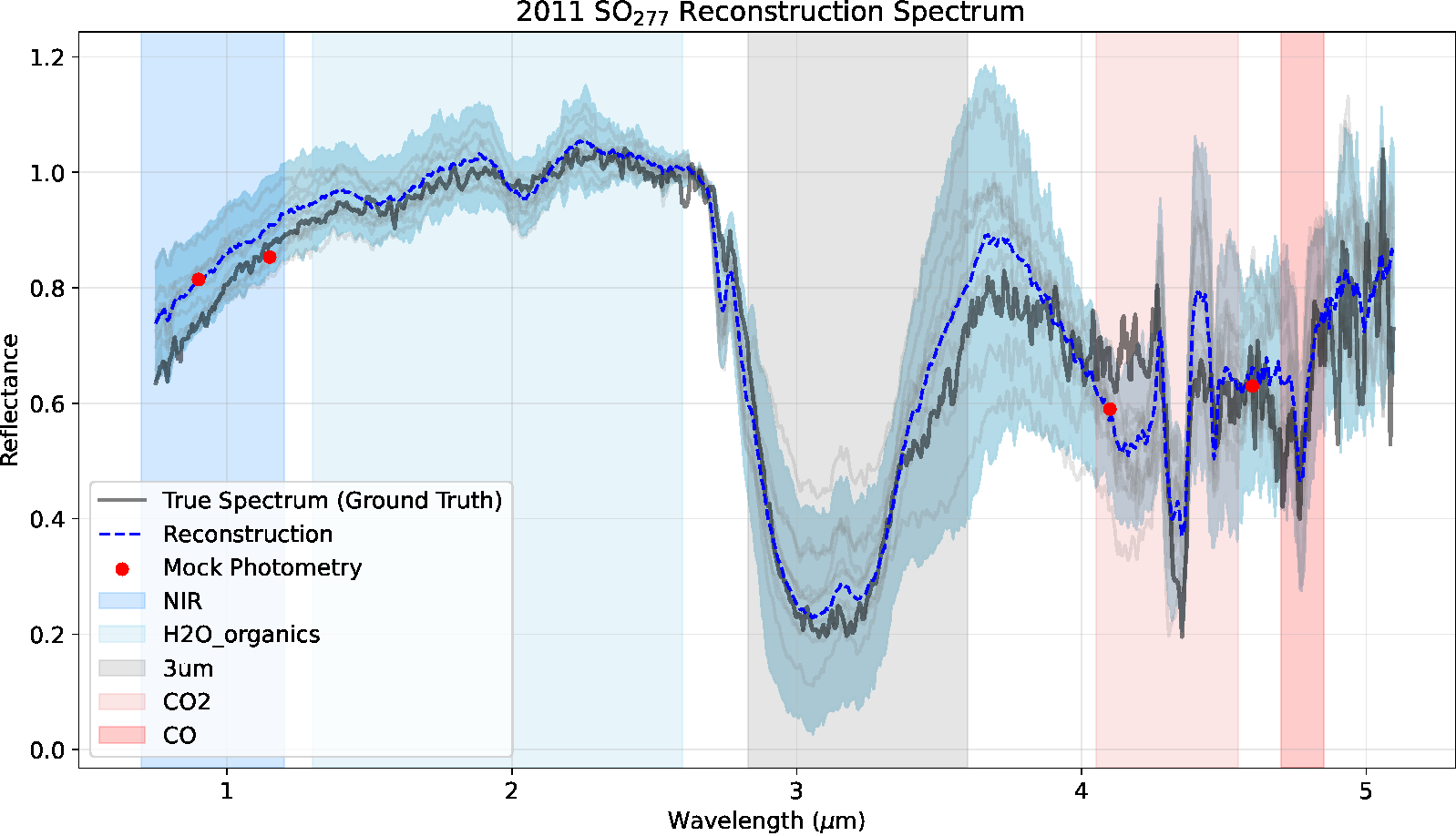}
\caption{Reconstructed spectra of the Neptune Trojan outliers 2006~RJ$_{103}$ and 2011~SO$_{277}$ derived from the 4-band filter configuration [F090W, F115W, F410M, F460M].}
\label{fig:outliers_4bands}
\end{figure}

Figure~\ref{fig:outliers_4bands} presents the spectral reconstructions of 2006~RJ$_{103}$ and 2011~SO$_{277}$ derived from the 4-band filter set [F090W, F115W, F410M, F460M]. Although the posterior mean spectra qualitatively resemble the overall manifold structure, the predictive uncertainty increases substantially relative to in-sample reconstructions. This widened credible interval reflects an appropriate expression of epistemic uncertainty under distribution shift; the ability to reduce confidence when operating in low-probability regions of the learned prior distinguishes this probabilistic framework from purely deterministic approaches.

\begin{figure}
\centering
\includegraphics[width=.7\columnwidth]{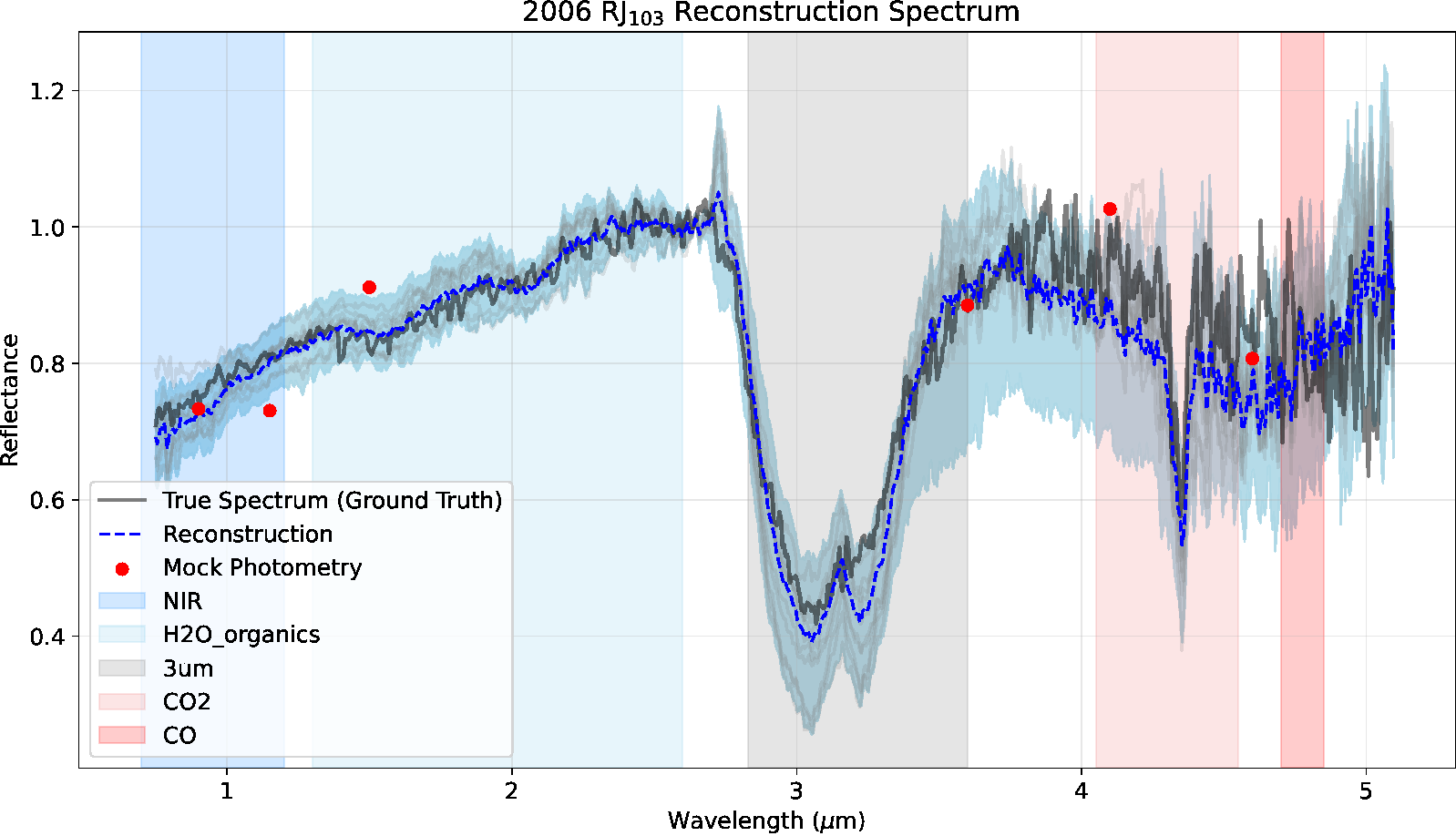}
\includegraphics[width=.7\columnwidth]{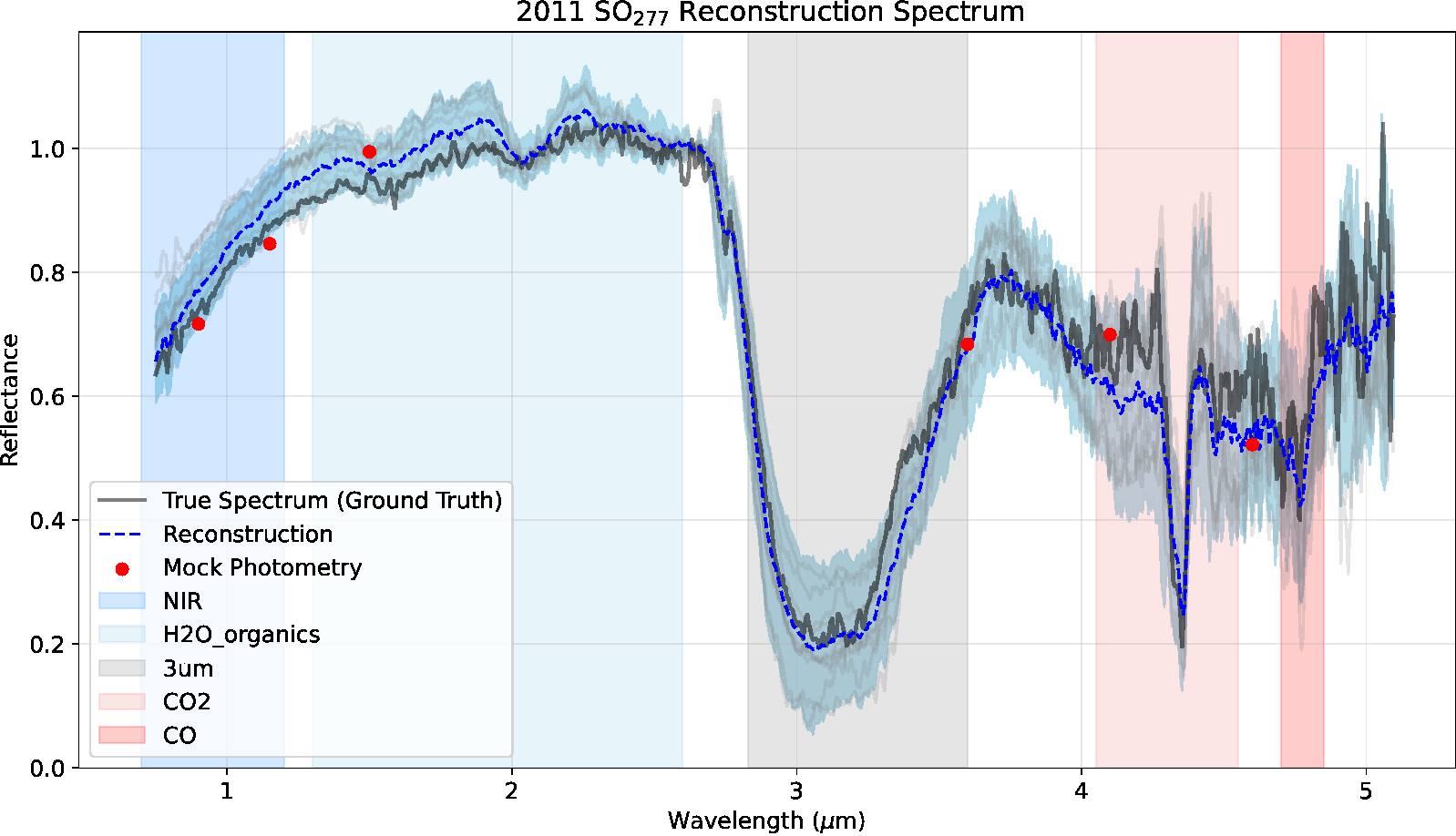}
\caption{Reconstruction spectra of 2006~RJ$_{103}$ and 2011~SO$_{277}$ with the 6-band photometry filter configuration [F090W, F115W, F150W, F360M, F410M, F460M]. The extra filters pinch the predictive reconstruction uncertainties.}
\label{fig:outliers_6bands}
\end{figure}

Figure~\ref{fig:outliers_6bands} illustrates the reconstructed spectra of the two Neptune Trojans using the 6-band photometry filter set [F090W, F115W, F150W, F360M, F410M, F460M]. With the object's specific location within the latent manifold more tightly constrained, the predictive uncertainty of the reconstruction shrinks as expected.

These experiments illustrate that the framework can highlight potential out-of-distribution morphologies through increased predictive uncertainty, although quantitative detection sensitivity depends strongly on filter configuration and training coverage. 
\textcolor{black}{For the current scale of near-infrared observations, manually identifying these signatures, such as inflated posterior variance or spatial divergence from the high-density prior regions is sufficient for anomaly detection. However, scaling this framework for automated discovery in upcoming massive photometric surveys (e.g., the Vera C. Rubin Observatory) will require formalizing these signatures into a strict, quantitative metric.}

\textcolor{black}{Future work will formalize this anomaly detection through the lens of information geometry. Rather than relying on simple point-to-point distances, we plan to utilize distribution-to-distribution metrics, such as the Kullback-Leibler (KL) divergence, to evaluate the information deviation between the posterior distribution of a newly observed target and the established multidimensional prior of the TNO manifold. By explicitly accounting for both the latent centroid positioning and the structural shape of the predictive covariance, such a metric will provide a rigorous, automated framework to flag out-of-distribution candidates for rapid spectroscopic follow-up.}

\section{Discussion} \label{sec:disc}
\subsection{Why the Spectral Reconstruction Works?}
\label{sec:why}
The performance of the spectral reconstruction pipeline may appear mathematically under-constrained, particularly when generating a full continuum ($0.7$ to $5~\mu$m) from as few as two photometric data points (e.g., Figure~\ref{fig:spec_2bands}). Here, we clarify the underlying statistical mechanism using the framework of Bayesian inference introduced in Equation~\ref{eq:bayesian_spectral}.

As established in Section~\ref{sec:method}, the entirety of our knowledge regarding TNO surface spectra is encoded in the \textbf{Prior}. Thus, the reconstructed spectra are strictly confined to a low-dimensional manifold; they cannot be arbitrary curves. They must conform to the covariance structures observed in the DiSCo-TNO training sample.

The reconstruction process can be visualized as a progressive collapse of probability space:

\begin{enumerate}
    \item \textbf{Zero Observations (The Prior):} Before any photometry is obtained, the system assumes the object could be any generic TNO. The probability distribution reflects the latent covariance structure learned from the training sample.
    
    \item \textbf{Single-Band Constraints:} If a single photometric point is obtained, it constrains the absolute flux (scaling) but provides no color information. The likelihood function is uniform across the shape parameter space; thus, the posterior remains equivalent to the Prior.
    
    \item \textbf{Two-Band Constraints (Slope Selection):} The addition of a second filter imposes a strong constraint on the allowed region of latent space. For example, a large color value for (F090W - F360M) immediately renders Water-type spectra (which are typically neutral/blue) highly unlikely. The likelihood function is no longer flat; it peaks sharply in the region of the Prior containing Red/Organic-type spectra. Consequently, the posterior probability for Water-type solutions collapses to near zero, even if the model has not "seen" the water absorption features directly.
    
    \item \textbf{Multi-Band Constraints (Feature Pinpointing):} As further observations are added (e.g., F115W and F460M), the previous posterior effectively becomes the new prior. We now possess constraints on both the Near-IR slope ($0.9-1.15~\mu$m) and the mid-IR absorption profile ($3.6-4.6~\mu$m). If the input vector shows a steep near-IR slope, the likelihood function strongly disfavors the shallower near-IR slope of CO$_2$-types, forcing the model to select Organic-type solutions. 
\end{enumerate}

In essence, the pipeline does not arbitrarily ``invent'' the spectrum between the filters; rather, it synthesizes the spectral shape by locating the coordinate within the learned continuous manifold that best satisfies the constraints imposed by the photometric likelihood.

\subsection{Why Some Bands Are More Informative Than Others?}
\label{sec:why2}
Our survey optimization analysis (Section~\ref{sec:bands}) yielded a clear hierarchy of filter importance, with short-wavelength (e.g., F090W) and long-wavelength (e.g., F460M) bands providing the most critical constraints. 

\textcolor{black}{To explain this result, we introduce the mathematical link between the PCA manifold and our survey optimization, grounded in the relationship between PCA loadings, variance, and information content. In the PCA framework, the feature loading at wavelength $\lambda$ for the $k$-th component is given by $L_k(\lambda) = v_k(\lambda)\sqrt{\sigma^2_k}$, where $v_k(\lambda)$ and $\sigma^2_k$ are the eigenvector and the eigenvalue (which represents the total variance of the component), respectively. Here, the squared loading $L_k^2(\lambda)$ represents the spectral density of variance. For a specific wavelength range, such as the band-pass of a filter $[\lambda_0, \lambda_1]$, the variance captured by that filter is:}
\begin{equation}
\label{eq:loading}
\int_{\lambda_{0}}^{\lambda_{1}} L_k^2(\lambda) d\lambda = \sigma^2_{k, \text{filter}}.
\end{equation}
\textcolor{black}{In the context of information theory, the continuous Shannon entropy $H_k$ for a Gaussian distribution associated with this captured variance is:}
\begin{equation}
\label{eq:entropy}
    H_{k, \text{filter}} = \frac{1}{2} \ln(2\pi e \sigma^2_{k, \text{filter}}).
\end{equation}
\textcolor{black}{Combining these concepts, we see that the prior Shannon entropy contained within a specific filter's band-pass for the $k$-th principal component is monotonically increasing with the captured variance ($\sigma^2_{k, \text{filter}}$) within that range.}

\begin{figure}
\centering
\includegraphics[width=.8\columnwidth]{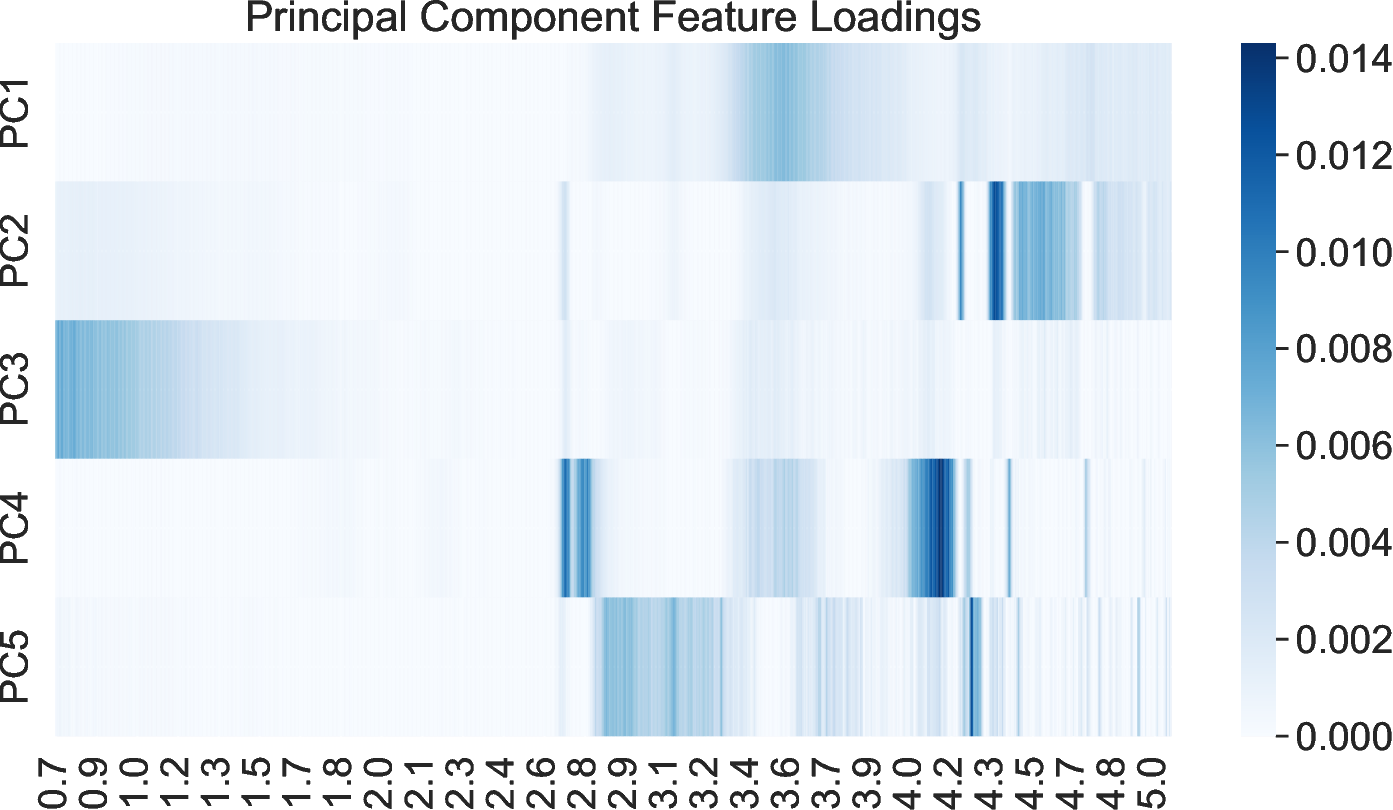}
\caption{The squared Principal Component Feature Loadings ($L_k^2(\lambda)$) of the TNO training spectra across the $0.7$--$5.0~\mu\text{m}$ range. We plot the squared loadings because $L_k^2(\lambda)$ mathematically represents the spectral density of variance for each principal component. As detailed in Section~\ref{sec:why2}, this variance is directly proportional to the reduction in Shannon entropy. Therefore, the color intensity serves as a map of information gain: deep blue regions indicate ``information hotspots" where photometric observations most effectively collapse the posterior uncertainty, while white regions ($L_k^2(\lambda) \approx 0$) represent negligible information content.}
\label{fig:loadings}
\end{figure}

\textcolor{black}{Figure~\ref{fig:loadings} visualizes the squared loading ($L_k^2(\lambda)$) of the training spectra. The blue regions indicate areas of high loading, high variance, and consequently, high prior entropy. The white areas represent low loading and low entropy. Since entropy quantifies the uncertainty or the number of possible states in a system, the white regions correspond to highly constrained spectral features with very little uncertainty; therefore, observations in these areas are not informative. On the other hand, the high-entropy (blue) regions contain significant uncertainty within the training data. Any photometric observation placed in these regions effectively collapses this uncertainty, yielding a massive information gain. This mathematically justifies why our identified optimal filters are so critical for spectral reconstruction.}

Looking at the feature loadings of our TNO training set in numerical order:
\begin{itemize}
    \item \textbf{PC1 (Continuum):} Shows broad loadings across the $3.0-4.0~\mu$m region, which are well-sampled by filters like \textbf{F356W} or \textbf{F360M}. This explains why optimal two-band configurations typically require pairing with one of these two filters, particularly when lacking further information regarding other PCs.

    \item \textbf{PC2 (CO$_2$ Absorption):} Features a high-amplitude loading region in the $4.3-4.7~\mu$m region, corresponding to the characteristic CO$_2$ absorption doublet and the CO absorption. The \textbf{F460M} filter overlaps with the shoulder of this feature. This explains why adding F460M drastically improves classification accuracy for CO$_2$-rich objects: it is the primary probe for the variance encoded in PC2.

    \item \textbf{PC3 (Optical Slope):} Exhibits its highest loadings in the short wavelength region ($\lambda < 1.0~\mu$m). The \textbf{F090W} filter is particularly informative because it directly samples the region of peak variance associated with PC3. Without F090W, the model cannot accurately constrain PC3, leading to degeneracy in the spectral slope determination.
    
    \item \textbf{PC4 (Shoulders of 3~$\mu$m Feature):} Features a high-amplitude loading region at $2.7~\mu$m and $4.1~\mu$m. This explains why F410M can be more informative than F430M in an optimized filter configuration. While F430M captures the high-amplitude $4.3~\mu$m region associated with PC2, it becomes redundant if F460M is already selected. In contrast, F410M captures the independent loading in PC4, rendering it more valuable than F430M in this context.
\end{itemize}

Therefore, the optimal filter sets typically involve F090W/F360M/F460M because each of them represents orthogonal sampling of the latent space: F090W constrains PC3, F360M constrains PC1, and F460M constrains PC2. The redundancy of intermediate filters (the filters with band-pass between $1.5$ to $2.7~\mu$m) occurs because the variance in those regions is either low (low loadings) or strongly covariant with the features already measured by the endpoints.

\textcolor{black}{We note a formal caveat regarding the Gaussian assumption utilized in Equation~\ref{eq:entropy}. The true distribution of TNO spectra in the latent space is complex and multimodal (see Figure~\ref{fig:latent}) due to the presence of distinct taxonomic classes, rather than strictly Gaussian. For non-Gaussian distributions, the true Shannon entropy may deviate quantitatively from the $\frac{1}{2}\ln(2\pi e \sigma^2)$ formulation. Nevertheless, the conceptual and qualitative framework remains robust: photometric filters with high captured variance ($\sigma^2_{k, \text{filter}}$) inherently correspond to regions of maximal structural diversity and uncertainty within the population. Thus, while the Gaussian entropy serves as an analytical proxy, the fundamental conclusion that the variance density ($L_k^2(\lambda)$) directly tracks the highest density of information content remains valid regardless of the exact underlying distribution.}

\subsection{Out-of-Distribution Behavior and Novel Surface Types}
\label{sec:ood}

While spectrally extreme populations such as Haumea-family members are easily identifiable through optical colors alone, more subtle out-of-distribution (OOD) cases arise when spectrally peculiar objects possess broadband optical colors that mimic the dominant TNO spectral classes. Rare surface types, including certain Neptune Trojans, can therefore remain indistinguishable in traditional color--color space despite possessing fundamentally different near-IR spectral architectures.

Within the present probabilistic framework, such objects do not require explicit prior labeling to be identified. Instead, they manifest through characteristic statistical signatures. Their posterior mean latent coordinates may reside in regions of low density under the kernel density estimate (KDE) prior, reflecting limited representation among the training spectra. Simultaneously, the posterior predictive variance becomes inflated relative to the LOOCV reference distribution, indicating reduced confidence in reconstruction within the learned covariance structure.

As demonstrated in Section~\ref{sec:outliers}, objects excluded from the training manifold are not reconstructed with artificially high confidence. Instead, the framework responds with broadened posterior credible intervals and reduced latent-space support. This elevated predictive uncertainty should not be interpreted as a model failure; it reflects the Bayesian formulation appropriately reverting toward the prior covariance in the absence of direct support from the empirical manifold. These signatures provide a data-driven mechanism for the probabilistic discovery of rare surface types embedded within color-degenerate populations, functioning entirely without the need for rigid parametric thresholds.

\subsection{Physical Implications of Low-Dimensional Spectral Structure}

The effectiveness of a low-dimensional latent representation suggests that the near-infrared spectral diversity within our sample is structured rather than purely stochastic. The primary latent components capture dominant trends in continuum slope and broad absorption morphology, while higher-order components refine localized features. This hierarchy indicates that the most significant modes of spectral variability can be described by a limited number of coordinated physical or chemical changes.

Surface spectra in the outer solar system are shaped by a complex interplay of ion and UV irradiation, volatile transport, thermal processing, and the mixing of ices and refractory organics \citep{Henault2025, DalleOre2015, Brunetto2006}. Laboratory studies demonstrate that irradiation of simple ices, such as H$_2$O, CH$_4$, CH$_3$OH, and CO$_2$, can modify both continuum slope and absorption band strengths in correlated ways \citep{Zhang2023, Cruikshank2019}. These maturation processes may therefore introduce intrinsic correlations among spectral features, such that variations in band depth and overall slope are not independent.

In this context, the success of our framework in constraining broad absorption morphology from sparse photometric measurements is consistent with the presence of coupled spectral variations driven by shared surface processes. The low intrinsic dimensionality inferred here suggests that physically realized surface compositions occupy a restricted region of the total available spectral space. This apparent spectral clustering may reflect a limited range of surface processing pathways operating within the TNO population. 

We note that low dimensionality in a linear PCA representation does not uniquely imply chemical simplicity; it may also reflect correlated spectral covariance within the observed sample. Further high-resolution spectroscopic observations, particularly from the James Webb Space Telescope (JWST), together with targeted laboratory experiments, will be necessary to map these latent dimensions onto specific chemical pathways and physical mechanisms.

\subsection{Assumptions and Limitations of the Framework}
\label{sec:limitations}

While this probabilistic framework successfully bridges the gap between sparse photometry and spectroscopy, its applicability is bounded by several inherent mathematical and observational limitations. It is critical to recognize these boundaries when deploying the pipeline on future datasets.

\begin{itemize}
    \item \textbf{Dependence on the Training Prior:} The most fundamental limitation of any generative model is its reliance on the training manifold. Because obtaining high signal-to-noise near-IR spectra of TNOs is observationally expensive, our PCA latent space is built upon a restricted seed population of $\sim$50 objects. Consequently, the prior represents the current observational understanding of the dominant TNO classes. As discussed in Section~\ref{sec:prior}, minority classes such as Haumea family members and Centaurs were deliberately excluded to maintain manifold stability. Therefore, the pipeline cannot accurately reconstruct the specific physical features of these excluded classes; it can only flag them as out-of-distribution outliers via inflated predictive uncertainty (as demonstrated in Section~\ref{sec:outliers}). More generally, any spectral morphology that lies outside the linear span of the learned eigenbasis cannot be faithfully reconstructed; such objects will instead manifest as poorly calibrated predictions or systematically inflated uncertainties.
    
    \item \textbf{The Linearity of PCA:} PCA is a linear dimensionality reduction technique. While this linearity successfully captures the broad continuum slopes and deep molecular absorptions of the major TNO classes, the underlying radiative transfer physics governing intimate mineral mixtures and space weathering is inherently non-linear. In regions of highly complex or blended molecular absorptions, a linear manifold may not fully capture non-linear mixing effects or localized absorption asymmetries, particularly if they are weakly represented in the training sample.
    
    \item \textbf{Simplified Noise Injection:} During the generation of the synthetic likelihood, we injected a uniform 20\% relative Gaussian uncertainty into the mock photometry to simulate a borderline $S/N \approx 5$ detection. In actual JWST/NIRCam observations, photometric uncertainties are highly wavelength-dependent, varying with instrument throughput, solar flux, and specific exposure times. While the flat 20\% noise acts as a robust, conservative regularizer for the AutoML regressor, applying this framework to a specific observational proposal would benefit from injecting the precise, instrument-specific noise model for those exact targets.
    
    \item \textbf{Unconstrained Wavelength Regimes:} As noted in the wavelength-dependent calibration analysis (Section~\ref{sec:wcal}), the accuracy of the reconstruction is ultimately bottlenecked by the physical bandpasses of the chosen filters. In regions lacking direct photometric coverage (e.g., the 2.7--2.9 $\mu$m gap between standard short and long-wavelength filters), the posterior relies entirely on the covariance structure of the prior. If an unseen object possesses a novel absorption feature strictly within an unobserved gap, the pipeline will remain blind to it. The framework therefore extrapolates according to the learned covariance structure rather than inferring genuinely new absorption behavior.

    \item \textbf{Binomial Uncertainty on Taxonomic Classification:} Although latent-space and photometric augmentation reduce variance in the supervised models, the effective prior remains constrained by the diversity of the original 51 seed spectra. Reported per-class classification accuracies therefore carry binomial uncertainty associated with the finite number of ground-truth objects. For the present sample sizes (6--22 objects per class), the corresponding 1$\sigma$ binomial uncertainty on per-class accuracy ranges from approximately $\pm$5--7\% for the largest class (Water and CO$_2$ -types) to $\pm$10--15\% for the smallest groups (the two Organic subclasses). These intervals reflect the limited size of the spectroscopic training set rather than any instability within the statistical framework itself. Within the observed manifold, class separability is strong and the supervised models are highly stable under LOOCV. Therefore, the dominant source of classification uncertainty arises from the finite sampling of the underlying TNO population. As additional high-quality near-IR spectra become available and the diversity of known surface types expands, the learned manifold and associated classification boundaries may evolve. The present survey optimization results are thus valid with respect to the current empirical taxonomy and will naturally refine as the training prior grows.

\end{itemize}

Acknowledging these limitations clarifies that the present framework should be interpreted as a probabilistic interpolation engine within the currently observed TNO spectral population, rather than a discovery tool for fundamentally new surface compositions. Its reliability scales directly with the breadth and diversity of the spectroscopic prior. As future high signal-to-noise surveys expand the catalog of ground-truth TNO spectra, the latent manifold can be correspondingly enlarged and, if warranted by sample size, extended to non-linear generative architectures. In this sense, the framework is inherently extensible and designed to evolve alongside the observational frontier.

\section{Summary \& Future Development}
\label{sec:summary}

\subsection{Summary}

In this work, we have developed and validated a probabilistic framework for reconstructing continuous near-infrared spectra of Trans-Neptunian Objects (TNOs) from sparse broadband photometry. By combining Principal Component Analysis (PCA) with Automated Machine Learning (AutoML), we map low-dimensional photometric observables onto a physically constrained latent manifold derived from high-quality spectroscopic priors.

Validation through leave-one-out cross-validation (LOOCV) demonstrates that the dominant modes of TNO spectral variability are effectively captured by a small number of latent components ($K \sim 5$), with posterior predictive intervals that are statistically well calibrated. Classification performance reflects the intrinsic geometric separability of surface types in latent space, achieving high per-class accuracies within the current empirical taxonomy. Residual uncertainties are driven primarily by finite sample statistics and filter coverage, rather than instability in the reconstruction framework itself.

While broadband photometry cannot reproduce fine-scale spectral structure, it retains sufficient information to probabilistically constrain broad absorption morphology and distinguish major taxonomic groups. Through survey optimization analysis, we quantified the information content of JWST/NIRCam filter combinations and identified a subset of near-infrared bands, F090W, F115W, F335M, F360M, F410M, and F460M, that most effectively resolve compositional degeneracies. 

Morever,  we introduced a probabilistic out-of-distribution (OOD) detection mechanism in which objects inconsistent with the learned spectral manifold are identified through inflated posterior predictive uncertainty and low prior density. This enables the discovery of rare surface morphologies, including dynamically distinct but color-degenerate populations such as peculiar Neptune Trojans, without requiring explicit prior labeling.

\textcolor{black}{Finally, as detailed in our discussion of the underlying statistical mechanism (Section~\ref{sec:why}), we emphasize that the success of this framework relies on the progressive collapse of the Bayesian probability space rather than arbitrary algorithmic extrapolation. Because the generative process is strictly confined to the pre-learned latent manifold (the Prior), even sparse constraints from just two or three photometric bands can sharply peak the likelihood function, effectively collapsing the posterior probability of unphysical or mismatched spectral shapes to near zero. This mathematically rigorous mechanism explains why sparse photometry is surprisingly powerful for full continuum reconstruction, confirming that our framework does not ``hallucinate'' missing data, but logically synthesizes it from established physical covariances. This makes it a highly reliable and interpretable tool for upcoming large-scale photometric surveys.}

\subsection{Future Outlook: Beyond Linear Manifolds}

The PCA-based framework presented here is intentionally conservative and well suited to the relatively smooth, continuum-dominated spectra of the current TNO sample. Given the limited number of high signal-to-noise near-infrared spectra presently available, a linear latent representation provides a stable and physically interpretable manifold.

As the catalog of ground-truth spectra expands, future iterations of this pipeline may transition to non-linear generative architectures capable of capturing more subtle compositional correlations. Transformer-based models, such as masked autoencoders \citep{he2021}, offer a promising direction by learning flexible spectral representations directly from incomplete photometric sequences without requiring explicit imputation. Such models naturally accommodate missing filters and heterogeneous survey strategies while preserving probabilistic reconstruction.

This generative philosophy is immediately scalable to minor planet populations with substantially larger spectroscopic training sets, such as main-belt asteroids (e.g., SMASS; \citealt{Bus2002a, Bus2002b, Xu1995}). In the era of the Vera C. Rubin Observatory, high-cadence multi-band photometry will be available for millions of small bodies. A generalized probabilistic reconstructor trained on extensive asteroid spectral libraries could transform broadband survey data into physically interpretable spectral classifications at unprecedented scale.

The framework introduced here therefore represents a first step toward probabilistic spectral inference across large photometric surveys, with dimensional complexity and model architecture naturally evolving alongside the growth of spectroscopic priors.

\section*{Software and Data Availability}
A demonstration of the generative spectral reconstruction pipeline, including a trained PCA manifold, is fully open-source and publicly available. The source code can be accessed on GitHub at \url{https://github.com/sevenlin123/spectra_reconstructor/}.

\begin{acknowledgments}

The authors are grateful to the anonymous referee for the thorough review and highly constructive feedback. The authors thank Noem\'{\i} Pinilla-Alonso and Rosario Brunetto for kindly providing us the reduced TNO spectra. The author (H.W.L.) is grateful to Zong-Fu Sie for insightful discussions that significantly enhanced the quality of this manuscript. All of the data presented in this paper were obtained from the Mikulski Archive for Space Telescopes (MAST) at the Space Telescope Science Institute. 
STScI is operated by the Association of Universities for Research in Astronomy, Inc., under NASA contract NAS5–26555. Support to MAST for these data is provided by the NASA Office of Space Science via grant NAG5–7584 and by other grants and contracts. 
Support for program \#2550 was provided by NASA through a grant from the Space Telescope Science Institute, which is operated by the Association of Universities for Research in Astronomy, Inc., under NASA contract NAS 5-03127. The authors acknowledge the support of the National Science Foundation under Grant No. AST‑2406527. 
L.M. was also supported by the Marsden Fund Council from Government funding, managed by Royal Society Te Apārangi. R.M. acknowledges support from NASA under Agreement No. 80NSSC21K0593 for the program “Alien Earths”.
\end{acknowledgments}

\vspace{5mm}
\facilities{JWST(NIRSpec)}


\software{AutoGluon \citep{agtabular}, astropy \citep{2013A&A...558A..33A,2018AJ....156..123A,astropy2022}, scipy \citep{2020SciPy-NMeth}, scikit-learn \citep{scikit-learn}, jwst \citep{2023zndo...8247246B}}

\bibliography{sample701}{}
\bibliographystyle{aasjournalv7}



\end{document}